\documentclass[11pt,a4paper]{article}
\pdfoutput=1

\usepackage{jheppub}

\usepackage{epsfig,psfrag,amsmath,amssymb,scalefnt}
\usepackage[dvipsnames]{xcolor}
\usepackage{amsmath,comment,braket}
\usepackage{placeins}
\usepackage{breqn}
\usepackage{slashed}
\usepackage{comment}
\usepackage{subcaption}
\usepackage{hyperref}
\usepackage[normalem]{ulem}

\allowdisplaybreaks

\preprint{
CERN-TH-2026-034,
P3H-26-017,
TTP26-006
}

\title{
  Analytic next-to-leading order electroweak
  corrections to Higgs boson pair production at high energies
}

\author[a]{Joshua Davies,}
\author[b]{Kay Sch\"onwald,}
\author[c]{Matthias Steinhauser,}
\author[b]{Hantian Zhang}

\affiliation[a]{Department of Mathematical Sciences, University of
    Liverpool, Liverpool, L69 3BX, UK}
\affiliation[b]{Theoretical Physics Department, CERN, 1211 Geneva 23, Switzerland}
\affiliation[c]{Institut f{\"u}r Theoretische Teilchenphysik, Karlsruhe Institute of Technology (KIT),  Wolfgang-Gaede Stra\ss{}e 1, 76131 Karlsruhe, Germany}

\abstract{
  We compute the complete top-quark-induced next-to-leading order electroweak corrections  to
  the form factors entering gluon-induced Higgs boson pair production.
  We consider the top quark contribution in the limit where the Mandelstam variables are much larger
  than all other scales involved in the process and compute about a hundred
  expansion terms in analytic form.
  They are used to obtain precise
  numerical results even for fairly low values of the transverse momentum of
  the Higgs boson.
  We show that these electroweak corrections at high energies are of the order of $-10\%$.
  We also discuss the leading logarithmic corrections of the analytic expressions.

}

\begin{document}
\maketitle
\flushbottom

%- {{{ Introduction:
\section{Introduction}

Higgs boson pair production is the most promising process to probe the Higgs self-coupling in the Standard Model~(SM) at the Large Hadron Collider~(LHC).
For this process, precise theoretical predictions %including higher-order corrections 
are indispensable to interpret the experimental data.
At the LHC, the dominant contribution to the cross section
comes from the loop-induced gluon-fusion process $gg\to HH$ that receives large higher-order corrections~\cite{Borowka:2016ehy,Borowka:2016ypz,Baglio:2018lrj,Davies:2019dfy}.

In this context, next-to-leading order~(NLO) two-loop electroweak~(EW) corrections to $gg \to HH$ at high energies are of both theoretical and phenomenological interest. 
The high-energy behaviour of scattering amplitudes beyond logarithmic and leading power approximations is a state-of-the-art problem in quantum field theory,
whose complexity further increases when the full electroweak sector of the SM is taken into account. 
The high-energy expansion of the
$gg\to HH$ amplitudes will be particularly useful for studying the production of boosted Higgs bosons at the LHC~\cite{ATLAS:2024lsk}, especially in the upcoming high-luminosity phase.

In the literature, there are a number of electroweak calculations available for $gg \to HH$.
The leading Yukawa corrections have been first considered in the high-energy expansion~\cite{Davies:2022ram} and the large top-mass limit~\cite{Muhlleitner:2022ijf}.
The first electroweak calculation involving top-quarks, gauge and Higgs bosons has been performed in a large top-mass expansion~\cite{Davies:2023npk}, where five expansion terms in $1/m_t^2$ are presented. 
The complete electroweak corrections including all SM contributions have been computed numerically~\cite{Bi:2023bnq}.
In addition, the contributions including Higgs self-couplings (and Yukawa) corrections have been computed with numerical approaches~\cite{Borowka:2018pxx,Li:2024iio,Heinrich:2024dnz} and analytic expansions~\cite{Bizon:2018syu,Davies:2025wke}.
Exact analytic results are available for the factorizable electroweak contributions~\cite{Zhang:2024rix} and the light-quark electroweak contributions~\cite{Bonetti:2025vfd}.
Combined results of Yukawa and light-quark contributions are also available in~\cite{Bhattacharya:2025egw,Bonetti:2025xtt}. Contributions from the quark-antiquark channel are available from Ref.~\cite{Bonetti:2026cih}.

In the QCD sector, the high-energy amplitudes have proven to be valuable for a number of phenomenologically important processes~\cite{Liu:2017vkm,Davies:2018ood,Davies:2018qvx,Catani:2022mfv,Wang:2023qbf,Jaskiewicz:2024xkd,Hu:2025aeo,Hu:2025hfc}.
For example, a deep high-energy expansion (with more than 100 expansion
terms) for $gg \to HH$~\cite{Davies:2018ood,Davies:2018qvx} has been successfully combined with the numerical approach~\cite{Davies:2019dfy} or forward-scattering expansions~\cite{Bonciani:2018omm,Bellafronte:2022jmo,Davies:2023vmj,Davies:2025qjr} to cover the full phase space for QCD corrections.
%%%%%

In this paper, we present the full analytic result for electroweak corrections in a deep high-energy expansion for $gg \to HH$ involving top quarks, gauge and Higgs bosons.
We will show that our result can cover a large phase-space region, ranging from the high-energy limit to fairly low values of the Higgs boson transverse momentum.
These results lead to
negative corrections of the order of $-10\%$ at the partonic level.

The outline of this paper is as follows: In Section~\ref{sec::conv} we present our conventions, and in Section~\ref{sec::tech}
we provide technical details on the computation of the Feynman diagrams and in particular on the
implementation of the expansions. 
In Section~\ref{sec::ren}
we describe the renormalization, for which we use the on-shell scheme. Results for the form factors are shown in Section~\ref{sec::res} where we also discuss their high-energy structures and describe our
approach to obtain Pad\'e-improved approximations.
We conclude in Section~\ref{sec::con}.
In the Appendix we provide explicit numerical results
for the one- and two-loop form factors for a few
selected phase-space points.

\section{Conventions}
\label{sec::conv}
We adopt the conventions of Section~2 of Ref.~\cite{Davies:2025wke}.
For convenience we repeat the most important
formulae in the following.

We consider the process $g(q_1)g(q_2)\to H(q_3)H(q_4)$, where all
momenta are incoming. The corresponding amplitude can be decomposed as
a linear combination of two Lorentz structures with scalar coefficients
${\cal M}_1$ and ${\cal M}_2$. The latter are in turn decomposed into
``box'' and ``triangle'' form factors as follows 
\begin{eqnarray}
  {\cal M}_1 &=& X_0 \, s \, 
  \left(F_{\rm tri1}
 + F_{\rm box1}\right)
                 \,,\nonumber\\
  {\cal M}_2 &=& X_0 \, s \, F_{\rm box2}
                 \,,
                 \label{eq::calM}
\end{eqnarray}
where we have introduced
\begin{align}
    F_{\rm tri1} & = \frac{3 m_H^2}{s-m_H^2}\Big(
          F_{\rm tri}
          + \frac{m_H^2}{s-m_H^2} \tilde{F}_{\rm tri} 
          \Big)\,.
\end{align}
$F_{\rm tri}$ and $\tilde{F}_{\rm tri}$ originate from Feynman diagrams with one and two cuts through $s$-channel Higgs boson propagators, respectively.
Note that at leading order (LO) we have $\tilde{F}_{\rm tri}=0$.
The prefactor in Eq.~(\ref{eq::calM}) is given by
\begin{eqnarray}
  X_0 &=& \frac{G_F}{\sqrt{2}} \frac{\alpha_s(\mu)}{4\pi}  \,,
\end{eqnarray}
where $\mu$ is the renormalization scale and $G_F$ is the
Fermi constant. 

For this process, the Mandelstam variables are given by
\begin{eqnarray}
  {s}=(q_1+q_2)^2\,,\qquad {t}=(q_1+q_3)^2\,,\qquad {u}=(q_2+q_3)^2\,,
  \label{eq::stu}
\end{eqnarray}
with
\begin{eqnarray}
  q_1^2=q_2^2=0\,,\qquad  q_3^2=q_4^2=m_H^{2}\,,\qquad
  {s}+{t}+{u}=2m_H^{2}\,.
  \label{eq::q_i^2}
\end{eqnarray}
One often either uses $\sqrt{s}$ and the transverse momentum of the Higgs
bosons in the center-of-mass frame, $p_T$, or $\sqrt{s}$ and the scattering
angle $\theta$ (also in the center-of-mass frame) in order to parametrize the
phase space. The relations to the Mandelstam variables are given by
\begin{eqnarray}
\label{eqn:thetadef}
  p_T^2 &=& \frac{{u}{t}-m_H^4}{{s}}\,,
          \nonumber\\
  t &=& m_H^2 - \frac{s}{2}\left(1-\cos\theta
        \,\sqrt{1-\frac{4m_H^2}{s}}\right)\,.
        \label{eq::pT_costhe}
\end{eqnarray}

We define the perturbative expansion of the form factors
as follows
\begin{eqnarray}
  F &=& F^{(0)} 
        + \frac{\alpha_s(\mu)}{\pi} F^{(1,0)} 
        + \frac{\alpha}{\pi} F^{(0,1)} 
        + \cdots
  \,,
  \label{eq::F}
\end{eqnarray}
where $\alpha_s$ is the strong coupling constant and $\alpha$ is the fine structure constant.
$F^{(1,0)}$ represents the two-loop
QCD corrections which are not discussed in this paper. 
%Their high-energy expansion has been computed in Ref.~\cite{Davies:2018ood,Davies:2018qvx}.

%- }}}
%- {{{ Technical details:

\section{Technical details}
\label{sec::tech}

\begin{figure}[t]
\centering
    \includegraphics[width=0.95\textwidth]{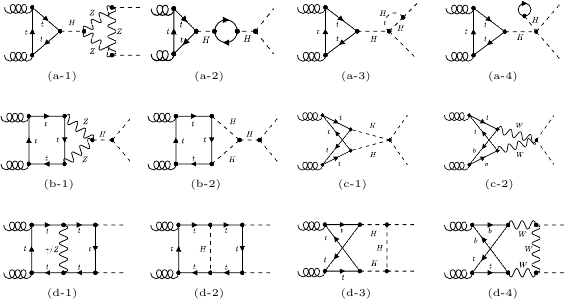}
  \caption{Sample Feynman diagrams contributing to
  the NLO electroweak corrections to $gg\to HH$.
  In the top, middle and bottom row, 1PR, triangle and box diagrams are shown. Note that (c-1) and (c-2) enter the box form factors since there is no cut through an $s$-channel Higgs boson propagator.}
   \label{fig::FeynDiag}
\end{figure}

Typical Feynman diagrams which contribute to the 
two-loop electroweak corrections are shown in Fig.~\ref{fig::FeynDiag}.
We want to stress that besides the one-particle irreducible (1PI)
contributions to the two-loop amplitudes there are
also one-particle reducible (1PR) contributions
(which have been computed in Ref.~\cite{Zhang:2024rix}). 
As a subset of the
latter we also include the so-called tadpole
contributions, which are necessary to obtain a
finite result (see Section~\ref{sec::ren}).

We consider all sectors of the SM.  However,
 since the light-quark contributions are known in analytic form from Ref.~\cite{Bonetti:2025vfd}
we concentrate on the top-quark-induced contributions. This means we 
consider the electroweak corrections generated by
diagrams with at least one top-quark propagator (not appearing in tadpole
contributions).
An exploratory study of the high-energy
behaviour for the diagrams with a Higgs boson exchange has been presented in
Ref.~\cite{Davies:2022ram}, where all fully-massive planar master integrals were
computed.  High-energy corrections for the subset of Feynman diagrams with non-zero top quark
Yukawa and Higgs self couplings were considered in Ref.~\cite{Davies:2025wke}.  In
this paper, deep analytic expansions for all fully-massive non-planar
master integrals have been obtained, 
and the results for the bare two-loop form factors have been
compared to the numerical results obtained in Ref.~\cite{Heinrich:2024dnz}.
In the region in which the expansions are valid, good agreement has been found.

A calculation of the complete NLO electroweak corrections 
with general gauge parameters has been performed in
Ref.~\cite{Davies:2023npk} in the large-$m_t$ limit, where it was demonstrated that the
gauge parameter dependence drops out for physical quantities.  
In the current 
paper we use the same set of diagrams but set all gauge parameters
($\gamma$, $Z$ and $W$) to Feynman gauge. As a consequence, the neutral and charged Goldstone bosons have the same masses as the corresponding
gauge bosons, $Z$ and $W$. A different choice of the gauge
parameters would introduce additional, but spurious, mass scales.

Our calculation involves several mass scales, which is currently too complicated to 
arrive at analytic results for the form factors.
For this reason we
perform several expansions: we perform Taylor expansions in the
final-state Higgs boson mass and in the mass differences of the top quark mass $m_t$
and the other boson masses in the loop diagrams. The resulting
integrals are expanded in the high-energy limit, where the Mandelstam
variables $s$ and $t$ are much larger than the top quark mass.

In the following we describe the computation
of the 1PI two-loop
diagrams. 
The Higgs boson mass appears in several places.
Since we process them differently we distinguish them in the
intermediate steps of our calculation.  We denote the mass of the
final-state Higgs bosons by $m_H^{\rm ext}$, the one in the
propagators inside loop diagrams by $m_H^{\rm int}$, and the Higgs
boson mass in the propagator of the triangle diagrams which is
outside of the loop by $m_H^{\rm prop}$.  Furthermore, in the SM the
Higgs boson mass appears in the expression for the Higgs boson self
coupling $\lambda$. We denote it by $m_H^\lambda$ thus have
\begin{eqnarray}
  \lambda &=& \frac{e^2 (m_H^\lambda)^2}{8 m_Z^2 c_{\rm w}^2 s_{\rm w}^2}
  \,,
\end{eqnarray}
where $e=\sqrt{4\pi\alpha}$, $\alpha$ is the fine structure constant, $c_{\rm w}\equiv\cos(\theta_{\rm w})$
is the cosine of the weak mixing angle, and $s_{\rm w}^2\equiv 1-c_{\rm w}^2$.

Note that for the box diagrams, $m_H^\lambda$ may have a different value
to the ``other'' Higgs boson masses
without affecting the renormalization of the amplitude. This is not the
case for the triangle contributions; here we must identify
$m_H^\lambda = m_H^{\rm prop}$ for a complete renormalization.
The values of $m_H^{\rm int}$ (which only appears at NLO) and $m_H^{\rm ext}$
(which is an external particle) can not interfere with the renormalization
procedure.
For numerical evaluation in Section \ref{sec::numresults}, we of course use the
same values for $m_H^{\rm ext}$, $m_H^{\rm int}$, $m_H^{\rm prop}$ and
$m_H^\lambda$, as given in Eq.~\ref{eq::numvals}.
   
Besides the Mandelstam variables $s$ and $t$, the additional scales are
the top quark mass $m_t$ and the gauge boson masses $m_Z$ and
$m_W$. For the latter we also distinguish the masses in the couplings
from those in the propagators inside loop diagrams, which we
denote by $m_Z^{\rm int}$ and $m_W^{\rm int}$, respectively.
Note that for any given diagram there are at most two different masses present:
$m_t$ and one of the boson masses $m_H^{\rm int}$, $m_Z^{\rm int}$ or
$m_W^{\rm int}$. Thus, for the generation of the form factors
expressed in terms of master integrals we proceed as follows.

  We use {\tt qgraf}~\cite{Nogueira:1991ex} for the generation of the amplitude,
  {\tt tapir}~\cite{Gerlach:2022qnc} for the translation to {\tt
  FORM}~\cite{Ruijl:2017dtg,Davies:2026cci} notation and the generation of auxiliary files
  for processing the integral topologies, and {\tt
    exp}~\cite{Harlander:1998cmq,Seidensticker:1999bb} for the mapping of each
  Feynman diagram onto a minimal subset of the integral topologies from {\tt
  tapir}.

  The masses of the internally propagating bosons are expanded around $m_t$ as a
  series in $\delta^\prime_X= 1 - (m_X^{\rm int}/m_t)^2$ with
  $X=H,Z,W$, see also Refs.~\cite{Davies:2022ram,Davies:2025wke}.  As
  a result, for the diagrams containing virtual $Z$ or Higgs bosons, all
  internal lines have the same mass: $m_t$.
  These integral families have already been classified in
  Refs.~\cite{Davies:2022ram,Davies:2025wke},
  where all relevant master
  integrals have been computed.
  For the diagrams with virtual $W$ bosons we have both massive lines (with
  mass $m_t$) and massless lines which originate from the bottom quark
  propagators. The integral families correspond to those
  of the QCD corrections have been defined in
  Refs.~\cite{Davies:2018ood,Davies:2018qvx}. Deep expansions of
  the master integrals have been computed in Ref.~\cite{Davies:2023vmj}.
  Further technical details regarding calculations of master integrals can be found in Refs.~\cite{Mishima:2018olh,Zhang:2024fcu}.
  In the final results we switch to the expansion parameter
  $\delta_X = 1 - m_X^{\rm int}/m_t$, since it shows better convergence
  properties (as e.g.~described in Ref.~\cite{Fael:2022frj}).
  We use the relation $\delta^\prime_X =
  \delta_X \left( 2 - \delta_X \right)$ and expand in $\delta_X$
  up to the same order which we had for~$\delta^\prime_X$.

 We compute the amplitude using the in-house {\tt FORM} code ``{\tt
    calc}'', which applies projectors to obtain the form factors, takes traces
  and expresses the final result in terms of scalar integrals of the integral
  families selected by {\tt exp}.  In this step we also perform the expansion
  of all boson propagators such that in the final results we have terms up to
  order $\delta_X^k$. For our calculation we have $k=3$, see also below. Note that there are no ``mixed'' terms in
  the $\delta_X$ expansion
  since Higgs, $Z$ and $W$ bosons never appear in the same Feynman diagram. At this point the scalar integrals depend on the
  variables $\{s,t,m_H^{\rm ext},m_t\}$; $\delta_X$ is only present in the
  integral coefficients.

   The next step is to expand in the final state Higgs boson mass which
  is encoded in the relation of the external momenta,
 the Mandelstam variables and $m_H^{\rm ext}$, see Eqs.~(\ref{eq::stu}) and~(\ref{eq::q_i^2}).
 We perform the expansion of each scalar integral
  at the level of the integrands with the help of
   {\tt
    LiteRed}~\cite{Lee:2012cn,Lee:2013mka}.  We generate {\tt FORM}
  {\tt Identify} statements which are applied to the summed expressions for
  the form factors. The remaining scalar integrals then depend only on
  $\{s,t,m_t\}$.
  The reduction to master integrals for the fully-massive families
  is performed with \texttt{Kira}~\cite{Maierhofer:2017gsa,Klappert:2020nbg,Lange:2025fba}.
  We extend
  the tables which have been generated for the calculation performed
  in Ref.~\cite{Davies:2025wke}. This part of the amplitude is
  expressed in terms of 168 master integrals, 28 of which are
  non-planar.
  For the QCD-like families we use the setup developed in
  Ref.~\cite{Davies:2018ood,Davies:2018qvx}. In particular, we use \texttt{FIRE}~\cite{Smirnov:2025prc} for the
  reduction and express the amplitude in terms of
  161 master integrals, 30 of which are non-planar.
  The reduction tables are inserted into the amplitude using \texttt{FORM}.
  This must be programmed carefully, to avoid both generating overly-large
  terms and blowing up the total number of terms in the expressions.
  
 Finally, we
  expand the amplitude in $\epsilon$ and for $m_t\to0$ and insert the
  analytic high-energy expansion of all master integrals from Refs.~\cite{Davies:2022ram,Davies:2018qvx,Davies:2018ood,Davies:2023vmj,Davies:2025wke}.
  This final stage of the computation is implemented efficiently
  in \texttt{FORM} by making use of Tablebases to store the $\epsilon$ and $m_t$
  expansion coefficients of the master integrals.

At this point we have obtained analytic expressions for the (bare) form factors
with explicit dependence on $s, t, m_t, m_H^{\rm ext}$ and $\delta_X$.
We have computed expansion terms up to $m_t^{108}$, $(m_H^{\rm ext})^4$
and $\delta_X^3$, and for $(m_H^{\rm ext})^0$ also $\delta_X^4$.
Due to the deep expansion in $m_t$ the expressions are quite large; they total
around 6.5~GB of gzip-compressed files.

In this work, we use \texttt{Mathematica} for the numerical evaluation at 200 digits or more,
which takes about half an hour
(with some dependence on the phase-space point).
For the calculation of hadronic cross sections, it is important to export these expressions to \texttt{Fortran} or \texttt{C} which will
lead to a significant reduction of the runtime. However, this must be done carefully so that the compilation time is no more than a few hours.

The next step is to renormalize the form factors, followed by their
numerical evaluation and Pad\'e approximation procedure. We discuss the
details of these parts of the calculation in the following sections.

%- }}}
%- {{{ Renormalization:

\section{\label{sec::ren}Renormalization}

For the NLO electroweak corrections there is no real-radiation
contribution and thus all $1/\epsilon$ poles are of ultraviolet
origin. They are removed after the renormalization
of an independent set of quantities. In our case we choose the
$W$, $Z$ and Higgs boson masses, the top quark mass and the
electric charge which is related to the fine structure constant
via $e = \sqrt{4\pi\alpha}$. We adopt the on-shell renormalization scheme
and introduce the $Z$ factors as follows
\begin{eqnarray}
  e^0 &=& Z_e e\,, \nonumber\\
  m_t^0 &=& Z_{m_t} m_t\,, \nonumber\\
  m_W^0 &=& Z_{m_W} m_W\,, \nonumber\\
  m_Z^0 &=& Z_{m_Z} m_Z\,, \nonumber\\
  m_H^0 &=& Z_{m_H} m_H\,,
  \label{eq::Z}
\end{eqnarray}
where the superscript ``0'' indicates bare quantities.  In addition to
the parameters in Eq.~(\ref{eq::Z}) we have to renormalize the wave
function of the Higgs boson
\begin{eqnarray}
  H^0 &=& \sqrt{Z_H} H\,,
  \label{eq::Z_H}
\end{eqnarray}
where $H$ is the Higgs boson field.  All $Z$ factors can be computed
from two-point functions; explicit analytic expressions can, e.g., be
found in Refs.~\cite{Denner:1991kt,Denner:2019vbn}.
For convenience we provide computer-readable
expressions in the ancillary files of this paper~\cite{progdata}.

Let us also stress that our NLO electroweak form factors do not have
an explicit dependence on the renormalization scale since all
parameters are renormalized in the on-shell scheme.

We consistently include tadpole contributions in all parts of our
calculation, i.e., in the bare two-loop amplitude and the $Z$ factors.
This prescription is equivalent to the so-called {\it
  Fleischer–Jegerlehner tadpole
  scheme}~\cite{Fleischer:1980ub}. Alternative options to treat
the tadpole contributions are discussed, e.g., in
Ref.~\cite{Dittmaier:2022maf}.

We renormalize the box and triangle contributions independently.  In
the case of the box contribution, at two-loop order there are only 1PI
Feynman diagrams (see Fig.~\ref{fig::FeynDiag}), 
and contributions form diagrams where one-loop tadpoles are attached to the (virtual)
top quark propagators. The identical tadpole contributions are present in the
on-shell renormalization constant $Z_{m_t}$.  Thus they 
exactly cancel in the on-shell scheme and finite results
for $F_{\rm box1}$ and $F_{\rm box2}$ are obtained from parameter and
wave function renormalization (as detailed above) by considering only
1PI contributions. 
This is different in the case of the triangle contribution where further
types of tadpole diagrams arise, see
Fig.~\ref{fig::FeynDiag}. Furthermore, there are various 1PR diagrams with
one-loop corrections to the Higgs boson propagator or one-loop vertex
corrections to the final-state triple-Higgs boson vertex.  We find it
convenient to sum all 1PI and 1PR contributions to the two-loop
amplitude and the renormalization constants and proceed with the
renormalization as described above. This leads to a finite results
for $F_{\rm tri1}$.

In the counterterm contribution we use for all (one-loop) $Z$ factors
exact (i.e.~non-expanded) results.  Since the $Z$ factors contain
$1/\epsilon$ poles we need the one-loop amplitudes including linear
terms in $\epsilon$. For the triangle diagrams the (exact) results can
be found in Ref.~\cite{Zhang:2024rix}. They contain at most
tri-logarithms. It is possible to compute the ${\cal O}(\epsilon)$ terms
of the box contributions, however the analytic expressions are
rather cumbersome. For this reason we use the deep high-energy expansion
of the one-loop box amplitudes from the calculation of the QCD corrections
in Ref.~\cite{Davies:2023vmj} (see also Ref.~\cite{Davies:2025wke}).

An important cross check of our result is the cancellation of the 
$1/\epsilon$ pole between the counterterm contribution and the two-loop amplitude.
If we perform a consistent expansion in $m_t$ we observe an exact cancellation.
Alternatively, if we leave the counterterm contribution unexpanded we typically
observe a cancellation of more than 100 significant digits in the coefficient of the
$1/\epsilon$ pole.
We have further verified numerically that our renormalized form factors are independent of explicit renormalization scale choices.

In a next step
we transform our results into
the so-called $G_\mu$ scheme where 
the fine structure constant $\alpha$ is replaced by
the Fermi constant $G_F$. 
For explicit expressions we refer, e.g.,  
to Section~5.1.1 of Ref.~\cite{Denner:2019vbn}.
The transformation to the $G_\mu$ scheme 
introduces the quantity $\Delta r$~\cite{Sirlin:1980nh}.
For an explicit result, see Eq.~(423)
of Ref.~\cite{Denner:2019vbn}.
In our numerical evaluation we do not
expand $\Delta r$ but evaluate it exactly,
and we evaluate the $Z$ factors similarly.
Note that $F^{(0,1)}$
in Eq.~(\ref{eq::F}) 
is already defined in the $G_\mu$ scheme.

%- }}}
%- {{{ Results:

\section{\label{sec::res}Results}

\subsection{Analytic results}

For convenience we provide explicit analytic results for the leading
non-vanishing terms of the LO
form factors. They are given by
\newcommand{\vph}{\vphantom{\Big\{}}
\newcommand{\lms}{l_{ms}}
\newcommand{\lts}{l_{ts}}
\newcommand{\lots}{l_{1ts}}
\begin{eqnarray}
  F^{(0)}_{\rm tri1}  &=&  \frac{6 m_t^2 m_H^2}{s^2} \biggl[4-\lms^2\biggr]
  + {\cal O}\left(\frac{m_t^4}{s^2}, \frac{m_H^4}{s^2}\right) \,,\nonumber\\
%%%
  F^{(0)}_{\rm box1} &=&  \frac{4m_t^2}{s}
  \biggl[2+\frac{m_H^2}{s}\biggl((\lots-\lts)^2+\pi^2\biggr)\biggr]
  + {\cal O}\left(\frac{m_t^4}{s^2},\frac{m_H^4}{s^2}\right) \,,\nonumber\\
%%%
        F^{(0)}_{\rm box2} &=&
                \frac{2 m_t^2}{s t (s+t)}\biggl[
                        -\lots^2 (s+t)^2
                        -\lts^2 t^2
                        -\pi^2 \left(s^2+2 s t+2 t^2\right)
                        + \frac{2 m_H^2}{s(s+t)} \biggl(
                                \lots^2 s (s+t)^2
                                \nonumber\\&&{}\vph
                                + \pi^2 s^3
                                + 2 s^2 t \left(-2 \lms+\lts+\pi^2-4\right)
                                - s t^2 \left(8 \lms+(\lts-2) \lts+16\right)
                                \nonumber\\&&{}\vph
                                - 4 (\lms+2) t^3
                        \biggr)
                \biggr]
        + {\cal O}\left(\frac{m_t^4}{s^2},\frac{m_H^4}{s^2}\right)\,,
        \label{eq::FF_LO}
\end{eqnarray}
where
\begin{eqnarray}
  \lms  &=& \log \left(\frac{m_t^2}{s}\right) +i\pi\,,\quad
  \lts  \,\,=\,\, \log\left(-\frac{t}{s}\right) +i\pi\,,\quad
  \lots \,\,=\,\, \log\left(1+\frac{t}{s}\right) +i\pi\,.
\end{eqnarray}

At two-loop order the form factors 
have a more complicated structure. However, 
the leading logarithmic contributions in $\log(m_t^2/s)$  are sufficiently compact. 
In the limit of vanishing external Higgs boson mass and small top-quark mass,
the leading logarithmic contributions in the high-energy limit for the box form factors are given by
\begin{align}
    F_{\rm box1}^{(0,1)} & = 
    \frac{m_W^4}{s^2  c_{\rm w}^6 s_{\rm w}^2} \Bigg[ \, l_{ms}^2 \,
    \frac{36 c_{\rm w}^6+32 c_{\rm w}^4-40 c_{\rm w}^2+17 }{36}  
    \Big(l_{1ts}^2+l_{ts}^2 -2 l_{1ts} l_{ts} +\pi ^2-4\Big)
    \Bigg] + \cdots,
    \label{eq::Fbox12l}\\[5pt]
    F_{\rm box2}^{(0,1)} & = 
    \frac{m_W^4}{s^2  c_{\rm w}^6 s_{\rm w}^2} \Bigg[ \, l_{ms}^3 \, \frac{36 c_{\rm w}^6+32 c_{\rm w}^4-40 c_{\rm w}^2+17}{27}   +\,  l_{ms}^2 \, \bigg(
     -\frac{2}{9}\, \delta_Z \, \left(32 c_{\rm w}^4-40 c_{\rm w}^2+17\right) \nonumber \\
     &\quad -8 \, \delta_W \, c_{\rm w}^6 +  
    \frac{36 c_{\rm w}^6+32 c_{\rm w}^4-40 c_{\rm w}^2+17 }{36 t (s+t)} \Big(l_{1ts}^2 (s+t)^2-2 \, l_{1ts} \, t (s+t)-2  \,l_{ts} \, t (s+t) \nonumber \\
    &\quad +\, l_{ts}^2 \, t^2 +\pi ^2 \left(s^2+2 s t+2 t^2\right)\Big)\bigg)
    \Bigg] + \cdots,
    \label{eq::Fbox22l}
\end{align}
where the ellipsis stand for sub-leading logarithmic contributions and higher-order terms in $m_t$, $m_H$ and $\delta_{W/Z} = 1-m_{W/Z}/m_t$. We note that the counterterms do
not contribute to the results 
shown 
in Eqs.~(\ref{eq::Ftri2l})--(\ref{eq::Fbox22l});
they come exclusively from the genuine two-loop diagrams. 
The leading term 
for the box form factors come from Feynman diagrams 
without Yukawa couplings as shown, e.g., in diagrams (d-3) and (d-4) in Fig.~\ref{fig::FeynDiag}. 
We obtain for the high energy limit of the triangle form factor
\begin{align}
    F_{\rm tri1}^{(0,1)} &=
     \frac{3 m_t^2 m_W^2}{16 s^2 s_{\rm w}^2} l_{ms}^4
    \left[ 
    -\frac{4 m_t^4}{m_W^4}
    +\frac{1 + 2 c_{\rm w}^4}{c_{\rm w}^4}  
    + \frac{4 s_{\rm w}^2 m_H^2}{27 c_{\rm w}^2 m_W^2}
    \right] + \cdots \,.
    \label{eq::Ftri2l}
\end{align}
We remark that while the high energy limits of the box form factors are obtained by expanding the corresponding two-loop diagrams directly, we obtain the high energy limit of the triangle form factor by re-expanding the partially exactly known results.
In the exactly known results, we do not keep track of external or internal Higgs boson masses.
We therefore obtain our result by strictly expanding in the $s \to \infty$ limit, but no re-expansion in any other scale is performed.
In order to shorten the expressions, we use 
$\log(m_X^2/s) + i \pi = l_{ms} + \mathcal{O}(\delta_X)$ and drop the terms that come from the $\delta$ expansion.

We observe that the leading term for $F_{\rm box1}$ is a quadratic logarithm,
whereas $F_{\rm box2}$ has a cubic contribution. Note that in the LO
expression shown in Eq.~(\ref{eq::FF_LO}) $F_{\rm box1}$ has no
$\log(m_t^2/s)$ terms at ${\cal O}(m_t^2)$ or ${\cal O}(m_H^2)$ and
the linear logarithms of $F_{\rm box2}$ are suppressed by
$m_H^2/s$. The LO triangle form factor has a quadratic $\log(m_t^2/s)$
contribution which is suppressed at high energies by a factor of $m_H^2/s$
due to the $s$-channel Higgs boson propagator.
It is also interesting to remark that $F_{\rm box2}^{(0,1)}$ has
$\log^2(m_t^2/s)$ terms which are proportional to $\delta_W$ or
$\delta_Z$.  
We note that at NLO EW, the process $gg \to HH$
contains several coupling structures, resulting in rich structures of logarithmic and
power corrections.
For further analytic expressions we
refer to the ancillary files~\cite{progdata}
where we provide five expansion terms
in the high-energy limit.
Deeper expansions are available on request.

The generic structure of the high-energy expansion of the 
form factors has the form
\begin{eqnarray}
  F &=& 
  \sum_{n=-4}^{108} 
  \sum_{i=0}^{2} \sum_{k_W=0}^{{4}}\sum_{k_Z=0}^{{4}}\sum_{k_H=0}^{{4}} 
        c_{ni}^{k_Wk_Zk_H} \: \left(\frac{m_t}{\sqrt{s}}\right)^{n} \:
        \left(\frac{m^{\rm ext}_H}{\sqrt{s}}\right)^{2i} \:
        \delta_W^{k_W} \: \delta_Z^{k_Z} \: \delta_H^{k_H} 
        \,,
        \label{eq::F_exp}
\end{eqnarray}
where negative powers in $m_t$ are only present for $i>0$.
Note that for $i=2$ we have $1/m_t^4$ terms.
Furthermore, there is the additional constraint that $k_W+k_Z+k_H \le 3$ 
for $i=1,2$ and $k_W+k_Z+k_H \le 4$ for $i=0$.
Since there are no ``mixed'' contributions in the 
$\delta_X$ expansions at most one of the indices
$k_W, k_Z, k_H$ may be non-zero.
For convenience we perform the rescaling
\begin{eqnarray}
  \delta_W^{k_W} &\to& \delta \,\,\delta_W^{k_W}\,, \nonumber\\
  \delta_Z^{k_Z} &\to& \delta \,\,\delta_Z^{k_Z}\,, \nonumber\\
  \delta_H^{k_H} &\to& \delta \,\,\delta_H^{k_H}\,,
\end{eqnarray}
and count in the following the powers of $\delta$.
The coefficients $c_{ni}^{k_Wk_Zk_H}$ in Eq.~(\ref{eq::F_exp})
depend on $t/s$ and $\log(m_t^2/s)$.

\subsection{Construction of approximations}
\label{sec::construction}

In the following, we describe our procedure to construct approximations. First,
we fix the maximal powers of $\delta$ and $(m_H^{\rm ext})^2$ and insert the
numerical values for $\delta_W$, $\delta_Z$, $\delta_H$ and $m_H^{\rm
  ext}$. We then specify the phase-space point
(i.e.~the values of $s$ and $t$).  This leads to an expansion in $m_t$. Next we insert
the numerical value for $m_t$ if it appears in the argument of
logarithms. Then we rescale odd and even power of $m_t$ according to
$m_t^{2k} \to m_t^{2k} x^k$ and $m_t^{2k-1} \to m_t^{2k-1} x^k$ and insert
also here the numerical value for $m_t$. As a result we obtain
a polynomial of degree $N=54$ in $x$. Afterwards we construct 
a Pad\'e approximant of the form
\begin{eqnarray}
  [n/m](x) &=& \frac{a_0 + a_1 x + \ldots + a_n x^n}{1 + b_1 x + \ldots + b_m x^m}
               \,,
\end{eqnarray}
where $n+m=N$. We can vary the parameters $n$ and $m$ and we can also choose
$N<54$. This provides the possibility to generate a large number of Pad\'e
approximants which allows us to determine a central value and uncertainty
interval.  This procedure is performed separately for each phase-space point.
More details can be found in Refs.~\cite{Davies:2020lpf,Davies:2019dfy}.

Whereas in the high-energy limit we can construct a deep expansion in $m_t$,
we only have a few terms in the expansion in $\delta$ and the final-state mass
$m_H^{\rm ext}$.  To improve convergence in these parameters,
sum acceleration techniques can be useful.  We use the so-called {\it Aitken
  extrapolation} (see, Ref.~\cite{Press:2007ipz}) and proceed as follows: The
starting point is an expansion of the form
\begin{align}
    A &= \sum\limits_{i=0}^{{\infty}} A_{i} \, y^i \,,
\end{align}
where in our case $y$ tags either the expansion in $\delta$ or in $(m_H^{\rm ext})^2$.
The $A_i$ correspond to the Pad\'e approximations based on the specified input and expansion depths, which are described below.
We then construct the partial sums
\begin{align}
    A^{(k)} &= \sum\limits_{i=0}^{k} A_{i} \, y^i ~,\qquad  k=0,1,\ldots,k_{\rm max}\,,
\end{align}
where for the expansion in $(m_H^{\rm ext})^2$ we have $k_{\max} =3$, 
while we have $k_{max}=4$ for the expansion in $\delta$.\footnote{{We set the expansion terms of $O(\delta^4 (m_H^{\rm ext})^i)$, $i>0$ to 0. The justification for this procedure is discussed in Sec.~\ref{sec::numresults}.}}
These partial sums are used to obtain the new approximation:
\begin{align}
    A_{\text{acc}}^{(n)} &= \frac{A^{(n)}A^{(n+2)} -
                           \left(A^{(n+1)}\right)^2}{A^{(n)} - 2 A^{(n+1)} +
                           A^{(n+2)}} \,. \label{eq::SumAcceleration}
\end{align}
It is interesting to note that $A_{\text{acc}}^{(n)}$ corresponds to the
$[n+1/1]$ Pad\'e approximant.

For the expansion in $\delta$ we can construct
the accelerated results $A_{\text{acc}}^{(0)}$, $A_{\text{acc}}^{(1)}$ and $A_{\text{acc}}^{(2)}$.
We can compare $A_{\text{acc}}^{(0)}$
and $A_{\text{acc}}^{(1)}$, which only include
expansions up to $\delta^2$ or $\delta^3$, 
respectively, to results which include $\delta^3$ or $\delta^4$ terms
without any acceleration.
We estimate the final
uncertainty of the $\delta$ expansion using the difference between our
best ``raw'' and accelerated results.

In principle we could also consider
sum acceleration in $m_H^{\rm ext}$.
However, as we will see below, the
expansion in $m_H^{\rm ext}$ converges sufficiently quickly that
the application of sum acceleration is not necessary.

\subsection{Numerical results}
\label{sec::numresults}

In this section we present numerical results
for the form factors and the squared partonic amplitude.
We use the following input values~\cite{PhysRevD.110.030001}
\begin{eqnarray}
\label{eq::numvals}
        G_F &=& 1.166378\times 10^{-5}\,\mbox{GeV}{{}^{-2}}\,,\nonumber\\
        m_H &=& 125.2\,\mbox{GeV}\,,\nonumber\\
        m_Z&=& 91.19\,\mbox{GeV}\,, \nonumber\\
        m_W &=& 80.37\,\mbox{GeV}\,, \nonumber\\
        m_t &=& 172.57\,\mbox{GeV}\,,
\end{eqnarray}
and introduce the coupling (e.g. in Eq.~(\ref{eq::F}))
\begin{eqnarray}
   \alpha_{G_\mu} &=& \frac{\sqrt{2}}{\pi} G_F m_W^2 (1-\frac{m_W^2}{m_Z^2})
   \,,
   \label{eq::alphaGmu}
\end{eqnarray}
which is our expansion parameter in the $G_\mu$ scheme.
The squared matrix element can be perturbatively expanded as
\begin{align}
    \mathcal{U} = \frac{1}{16} \left( |\mathcal{M}_1|^2 + |\mathcal{M}_2|^2 \right) = \frac{(X_0 s)^2}{16} \left( \mathcal{U}^{(0)} + \frac{\alpha_{s}(\mu)}{\pi} \,\mathcal{U}^{(1,0)} + \frac{\alpha}{\pi} \,\mathcal{U}^{(0,1)} + \cdots \right)\,,
\end{align}
which we use to define the ratio
\begin{align}
    r_{\rm EW} = \frac{\alpha}{\pi} \, \frac{\mathcal{U}^{(0,1)}}{\mathcal{U}^{(0)}}\,,
    \label{eq::rEW}
\end{align}
which quantifies the size of the NLO corrections at the partonic level.
In the results discussed below we use for the
LO term in the denominator of Eq.~(\ref{eq::rEW})
the expanded expressions. For all phase-space points which are considered in this paper this
provides an excellent approximation to the exact results.

\begin{figure}[t]
  \begin{tabular}{cc}
    \includegraphics[width=0.45\textwidth]{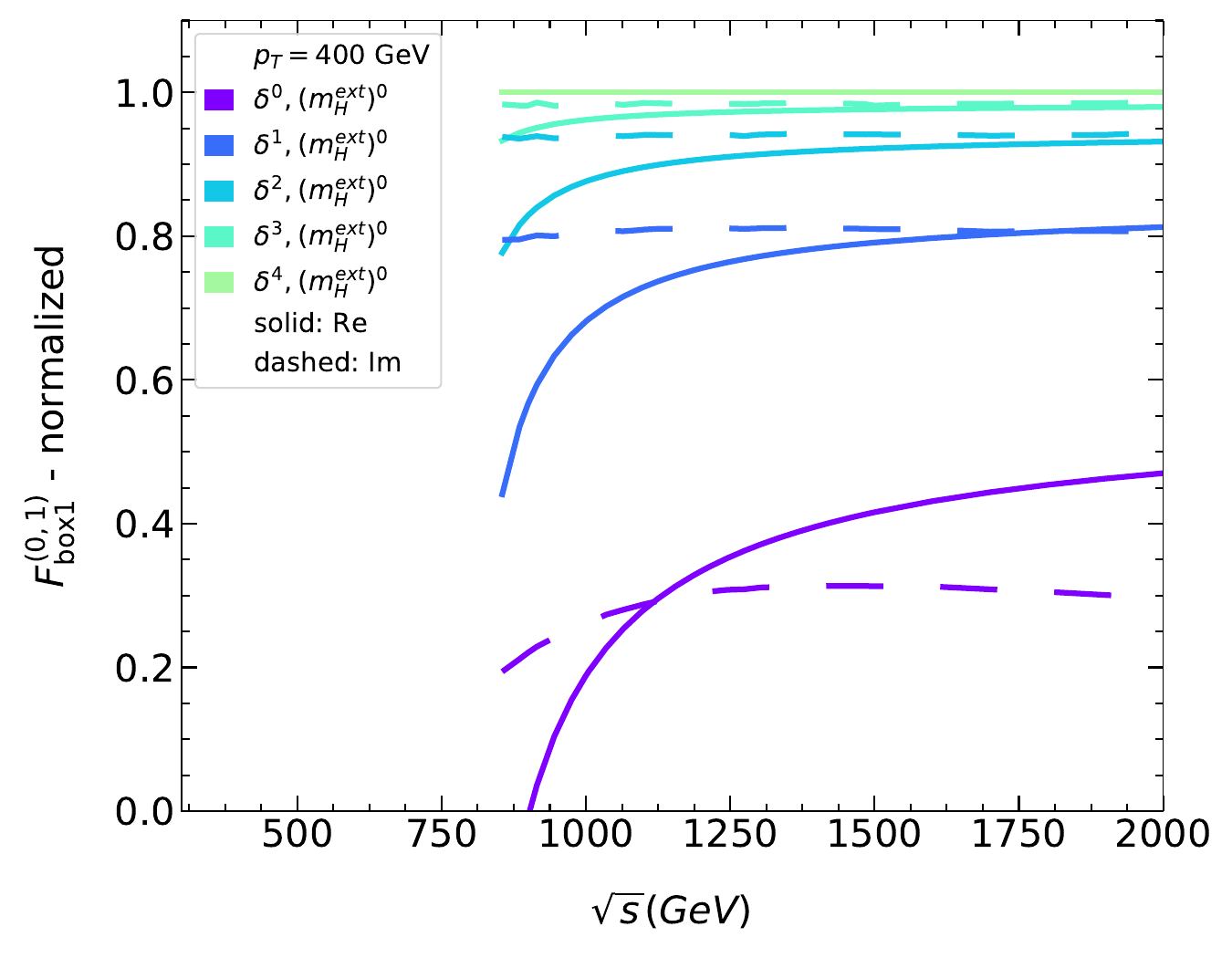}
    &
    \includegraphics[width=0.45\textwidth]{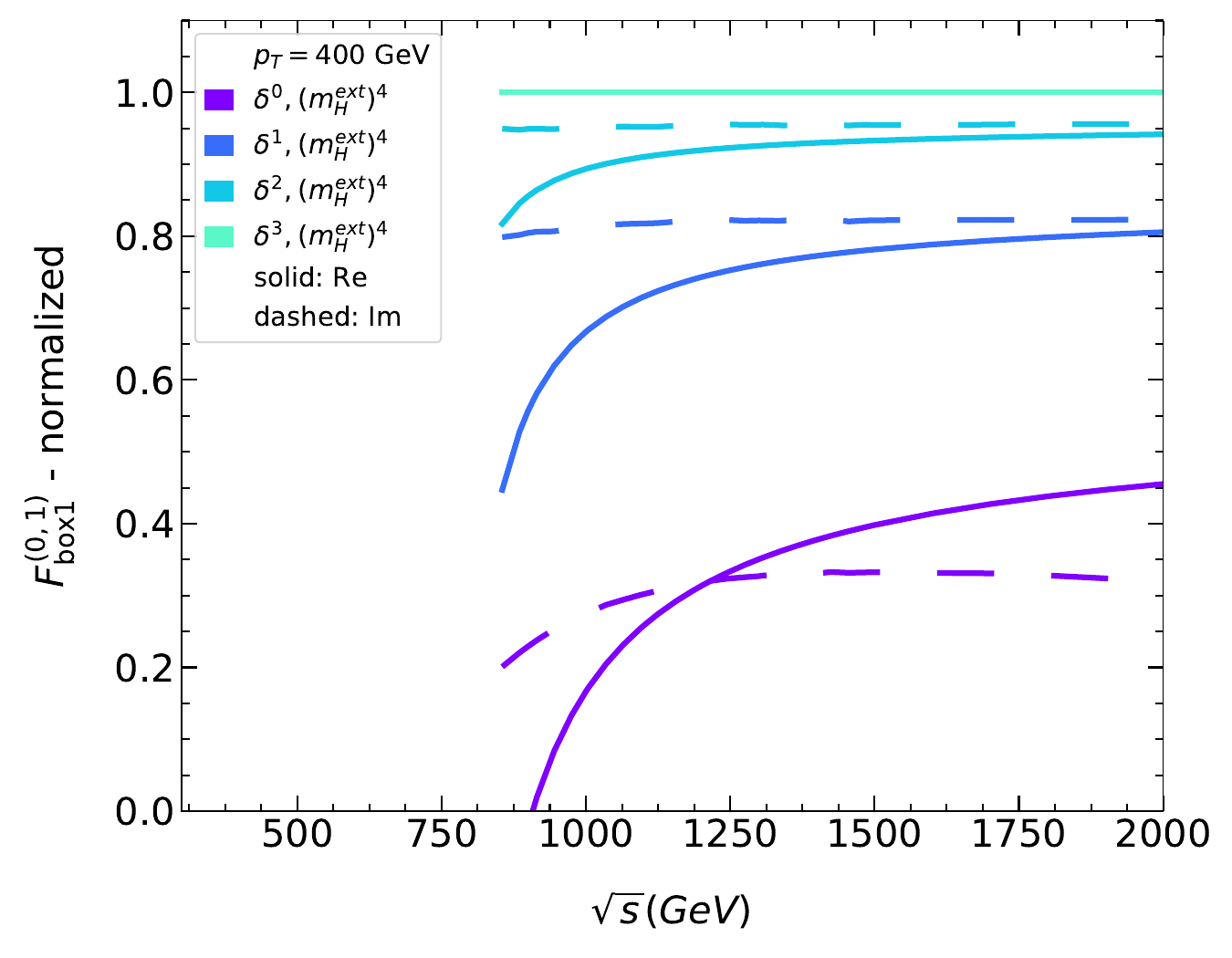}
    \\
    \includegraphics[width=0.45\textwidth]{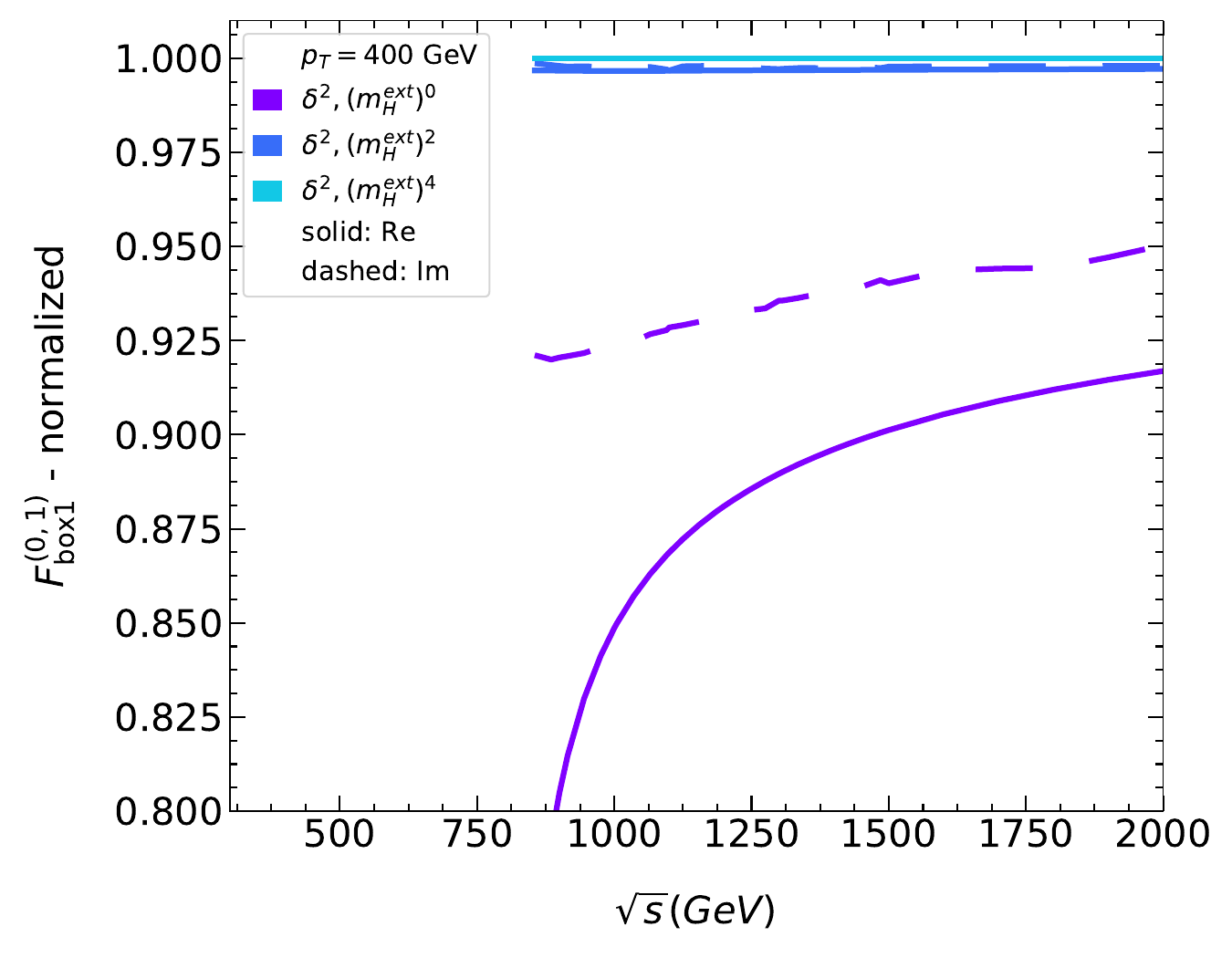}
    &
    \includegraphics[width=0.45\textwidth]{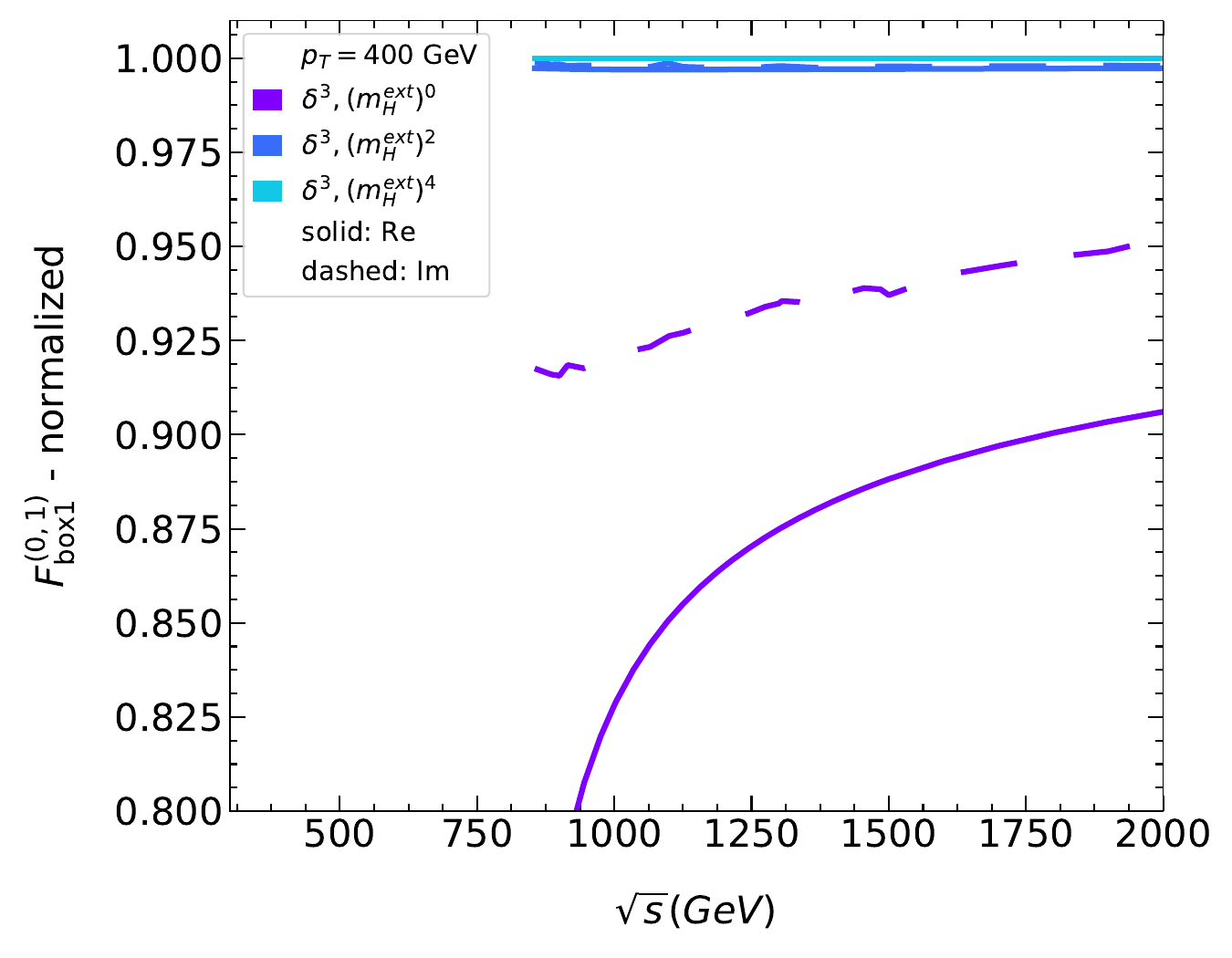}
  \end{tabular}
  \caption{\label{fig::FF_he_ra}
   Ratio of various $\delta$ and $m_H$ expansion terms normalized to the
   highest available approximation
   for $F_{\rm box1}$.
   Results for the real and imaginary parts are shown 
   as solid and dashed lines, respectively.
   The expansion depths given in the legend
   indicate the expansion terms which
   are used for the construction of the Pad\'e approximation. Here ``$\delta^n, (m_H^{\rm ext})^m$'' means that all expansion terms  $\delta^k$ and $(m_H^{\rm ext})^l$ with $k\le n$ and $l\le m $ are included.
   Top row: Fixed expansion depth in $m_H$. Bottom row: Fixed expansion depth in~$\delta$.
   }
\end{figure}

\begin{figure}[t]
\centering
\begin{tabular}{c}
    \includegraphics[width=.75\textwidth]{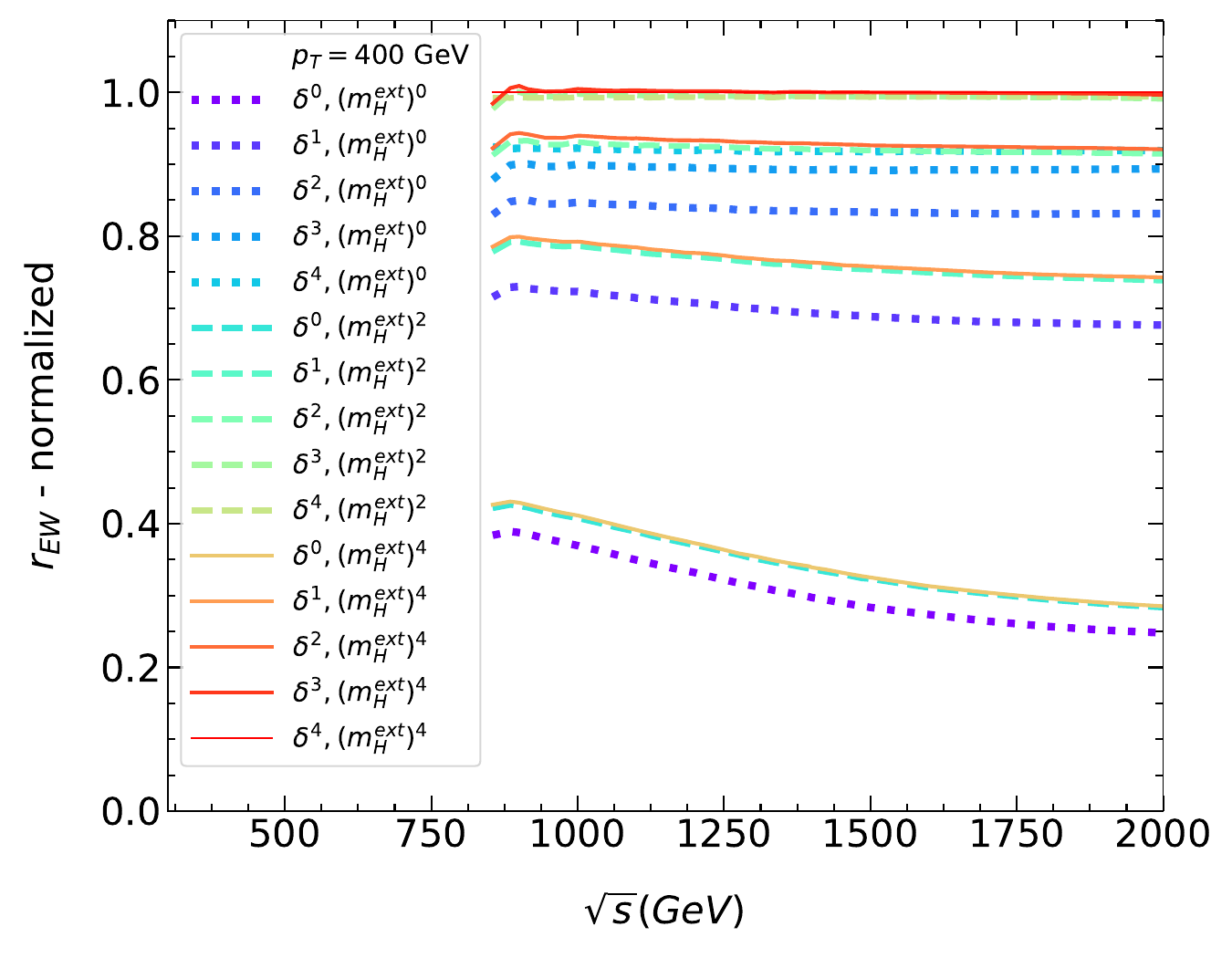}
\end{tabular}
  \caption{\label{fig::FF_he_ME_ra}
   $r_{\rm EW}$ for various $\delta$ and $m_H$ expansion terms normalized to
   highest available approximation for  $p_T=400$~GeV. For details concerning the
   meaning of the legend we refer to the caption
   of Fig.~\ref{fig::FF_he_ra}.
   }
\end{figure}

In the following, we describe Pad\'e results which have been obtained using
expansion depths between   $(m_t^2)^{46}$ and 
$(m_t^2)^{54}$.
Furthermore, we include 
approximations up to 
($m_H^{\rm ext})^4$
and $\delta^4$.
Coefficients which are not available
(i.e.~$(m_H^{\rm ext})^2 \delta^4$ and $(m_H^{\rm ext})^4\delta^4$)
are set to zero. Their numerical effect is expected to be small
since the quartic $m_H^{\rm ext}$ terms are, in general, of the order of $1\%$ or smaller, see below for more details.

Our form factors are expressed in terms of the Mandelstam variables $s$, $t$ and the masses for the top quark, gauge and Higgs bosons. For the numerical results discussed below, we
use $p_T$ and $\sqrt{s}$ as input and compute the value of $t$ using
\begin{align}
    t = \frac{1}{2}\left( 2m_H^2 - s + \sqrt{s}\sqrt{s-4 m_H^2 - 4 p_T^2} \right) \,,
    \label{eq::t}
\end{align}
which is then inserted in the analytic expression for the form factors. We hold one of the
      parameters $\sqrt{s}$ or $p_T$ fixed and examine the dependence on the other. For a
      fixed transverse momentum, we choose the minimum value for
      $\sqrt{s}$ to be either the value dictated by the kinematic
      constraints or approximately $2m_t$.
      In the Appendix we provide explicit numerical results
for the one- and two-loop form factors for a few
selected phase-space points.

In Figs.~\ref{fig::FF_he_ra} and~\ref{fig::FF_he_ME_ra} we discuss the
quality of the $\delta$ and $m_H^{\rm ext}$ expansions.  In
Fig.~\ref{fig::FF_he_ra} we show results for $F^{(0,1)}_{\rm box1}$
for $p_T=400$~GeV. In the different panels we either vary the
expansion depth in $\delta$ (top row) or $m_H^{\rm ext}$ (bottom row)
and normalize the results in each case to the highest available
approximation.  We observe a good convergence of the $\delta$
expansion; 
in each step the distance to the next expansion order is reduced by
more than a factor of two. 
Looking at the  $(m_H^{\rm ext})^0$ terms, where up to $\delta^4$ terms are available, we observe a rather flat correction that induces a shift of approximately $7.5\%$ for the $\delta^3$ term, and 
approximately $3\%$ for the $\delta^4$ term.
The missing term of order $\delta^5 (m_H^{\rm ext})^0$ can therefore be estimated to be around $1\%$--$1.5\%$.
A similar convergence pattern is observed for $F^{(0,1)}_{\rm box2}$.

The convergence in $m_H^{\rm ext}$ is even faster,
as can be seen from the panels in the bottom row
of Fig.~\ref{fig::FF_he_ra}. Here we conclude that 
the quadratic terms in $m_H^{\rm ext}$ are of the order of
$10\%$ and the quartic terms are at the $1\%$ level or below.
The size of the missing terms in the $m_H^{\rm ext}$ expansion are therefore negligible compared to the missing terms in the $\delta$ expansion.

From the observed convergence pattern we can estimate the size of the
missing $m_H^{\rm ext}$ terms at quartic order in $\delta$. 
For the
$\delta^4 (m_H^{\rm ext})^2$ $[\delta^4 (m_H^{\rm ext})^4]$ terms we
obtain $3\%\cdot10\% = 0.3\%$ [$3\%\cdot 1\% = 0.03\%$]
as compared to the known expansion terms. In our numerical analysis,
we set these expansion terms to zero (see Section~\ref{sec::tech} for details) at the point where
we still use $\delta^\prime$ (i.e.~quadratic mass ratio) for the parametrization of the expansion in mass differences. Note that
converting to the variable $\delta$ (i.e.~linear mass ratio) introduces terms up to order $\delta^4$, which we keep.
As we have shown above, dropping or keeping these terms only amounts to a small shift in our final numerical values.
In order to account for the finite expansion orders of our result, we assign 
a conservative uncertainty estimate of $\pm 2\%$.

In Fig.~\ref{fig::FF_he_ME_ra} we test our assumption  for
$r_{\rm EW}$. We
choose again $p_T=400$~GeV.  The dotted (blueish), dashed (greenish)
and solid (reddish) curves show the $\delta$ expansion in case no,
quadratic, or quartic $m_H^{\rm ext}$ terms are included in the
computation of $r_{\rm EW}$. In each case we construct five
approximations which differ in the expansion depth in $\delta$.  In
total 15 different approximations are shown in
Fig.~\ref{fig::FF_he_ME_ra}. All results are normalized to the
approximation which includes $\delta^4 (m_H^{\rm ext})^4$
terms.\footnote{Note that this approximation
contains all expansion terms $\delta^n$ and $(m_H^{ext})^m$
with $n\le3$ and $m\le4$, and the $\delta^4 (m_H^{ext})^0$
term. However, the (numerically small) expansion terms 
$\delta^4 (m_H^{\rm ext})^2$ and $\delta^4 (m_H^{\rm ext})^4$ are set 
to zero, as described above.}  We observe a similar convergence
pattern as in Fig.~\ref{fig::FF_he_ra}. In particular, the dashed and
solid curves, which correspond to the same $\delta$ expansion, almost
lie on top of each other indicating that the quartic $m_H^{\rm ext}$
terms are below 1\%.  Furthermore, the curves which belong to
$\delta^3 (m_H^{\rm ext})^2$, $\delta^4 (m_H^{\rm ext})^2$ and $\delta^3
(m_H^{\rm ext})^4$ are very close to reference line at ``$1$'',
which supports our estimate that the higher order expansion terms in
$\delta$ and $m_H^{\rm ext}$ are below a $\pm 1\%$ correction.

We observe a very similar convergence pattern
for all values of $p_T\gtrsim 300$~GeV.
This suggests the assignment of a total uncertainty of $\pm1\%$
to the squared NLO matrix element when it is used as input for the hadronic cross sections.
This estimate is further backed by the sum acceleration introduced in Section~\ref{sec::construction}.
We use sum acceleration for the expansion in $\delta$, since it shows the slower overall convergence.
We observe that $A_{\text{acc}}^{(0)}$, where no $\delta^3$ terms enter, reproduces our nominal result including $\delta^3$ terms within about $5\%$. Similarly, $A_{\text{acc}}^{(1)}$
reproduces the nominal result including $\delta^4$ terms to order $1\%$.
Furthermore, 
we observe that the  difference between our nominal best approximation
and $A_{\text{acc}}^{(2)}$ is below 1\%,
which again supports the conservative uncertainty estimate
of 1\% for $r_{\rm EW}$.

%\clearpage

\begin{figure}[t]
  \begin{tabular}{cc}
   \includegraphics[width=.45\textwidth]{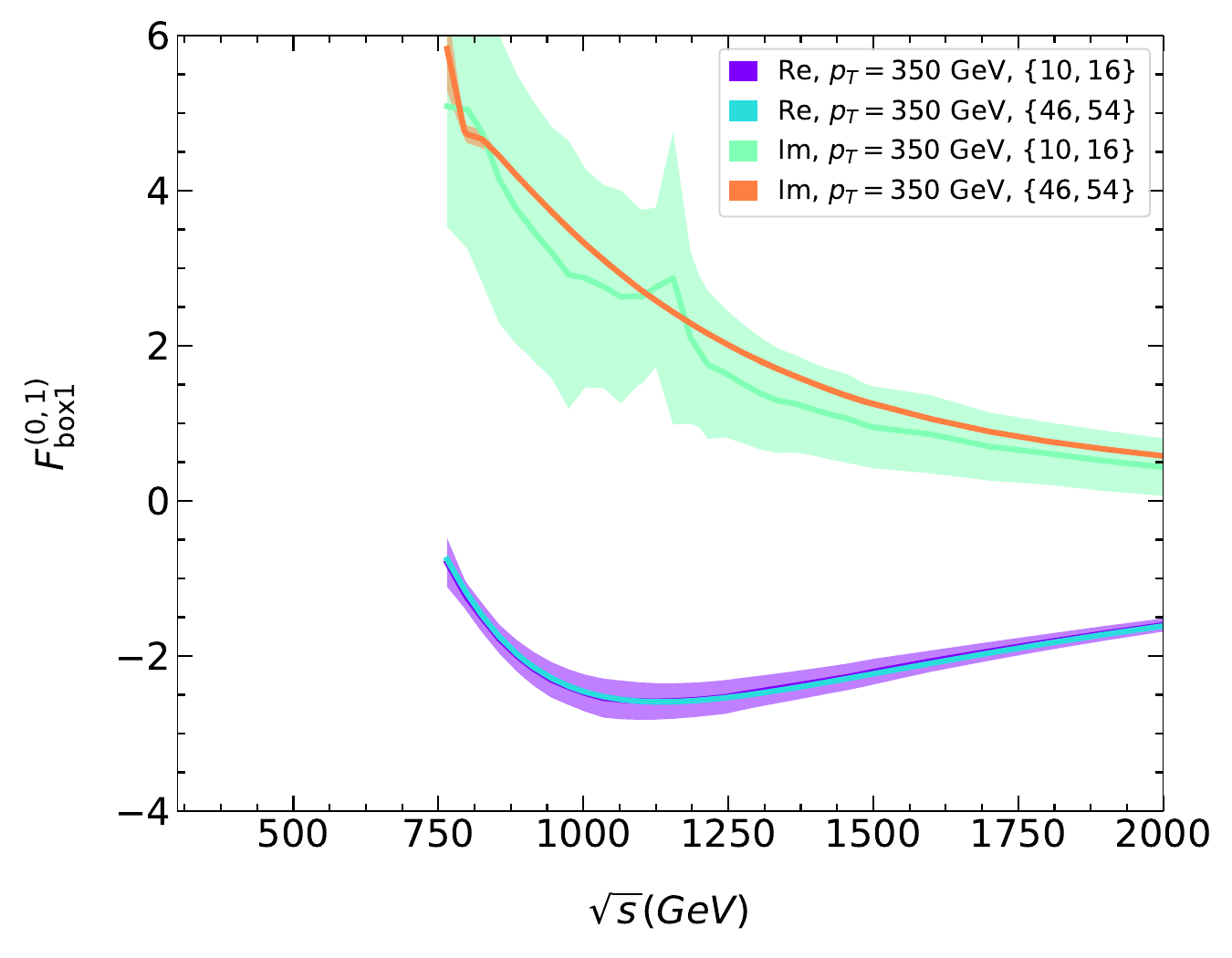}
    &
   \includegraphics[width=.45\textwidth]{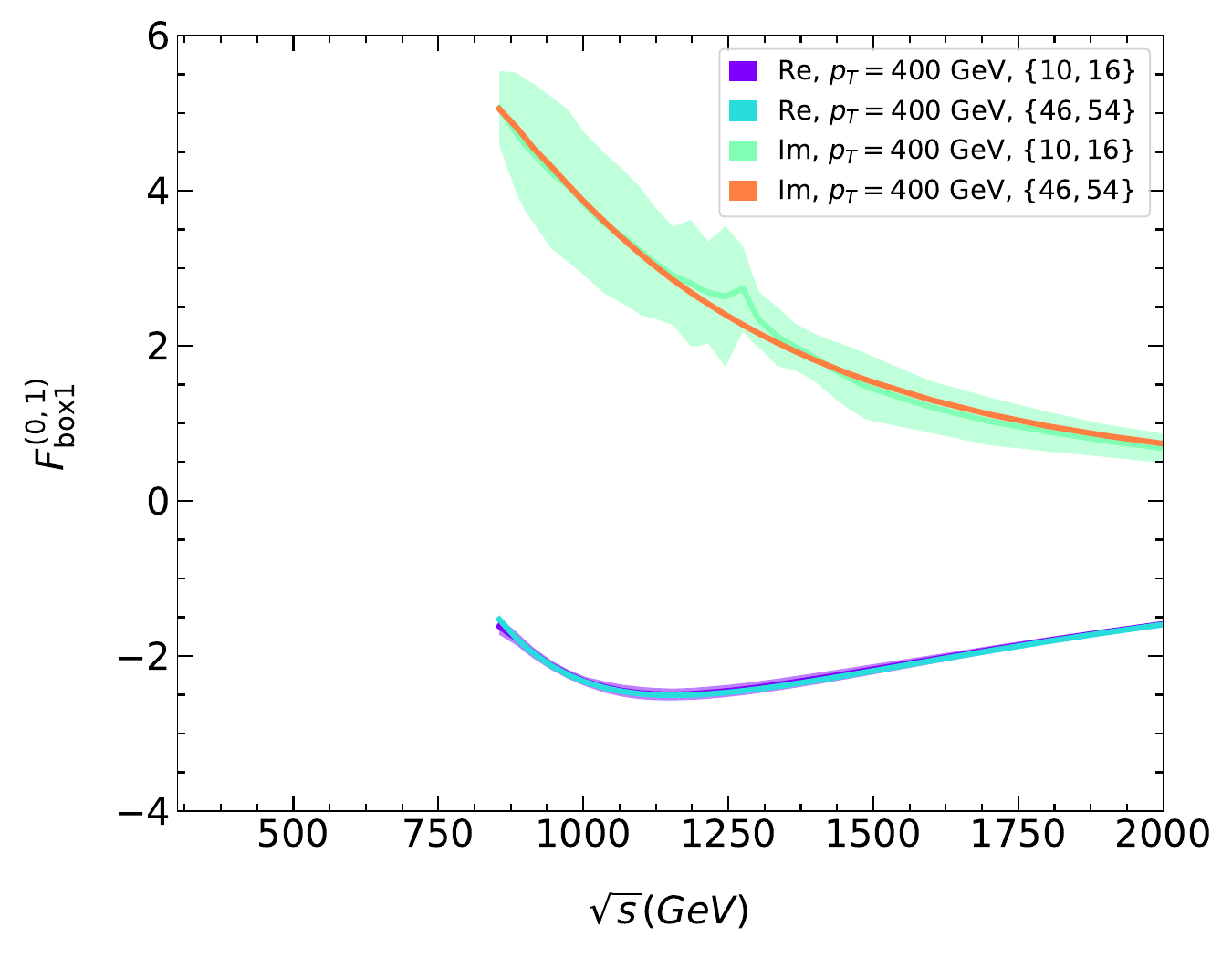}
  \end{tabular}
  \caption{\label{fig::FF_mt}Real and imaginary parts of 
   $F_{\rm box1}^{(0,1)}$ for $p_T=350$~GeV and $p_T=400$~GeV as a function of $\sqrt{s}$ for different expansion depth in $m_t$ used
   for the Pad\'e approximation, as indicated in the legend:
   $\{n,m\}$ means that for the construction of the Pad\'e 
   approximation at most expansions up to $(m_t^2)^m$ 
   and at least up to 
   $(m_t^2)^n$  are taken into account.}
\end{figure}

Next we discuss the dependence on expansion depth in $m_t$.
In Fig.~\ref{fig::FF_mt} 
we show the real and imaginary parts of
$F_{\rm box1}^{(0,1)}$ for two different values of $p_T$, as a function of $\sqrt{s}$. 
In each case we show two curves
and their associated uncertainty bands, which are
all based on expansions
up to $\delta^4$ and $m_H^2$ but differ in the
highest power in $m_t$ used for the construction of the
Pad\'e-improved result, as indicated in the legend.
We observe sizeable uncertainty bands for the 
approximations containing at most $m_t^{32}$
terms, in particular for the imaginary part.
The uncertainty is reduced substantially once 
we allow terms up to $m_t^{108}$ terms.
The latter is completely contained within the uncertainty band of the
lowest approximation.

Here we note that for some phase-space points, which correspond
      to certain values of $t$, the $m_t$ expansion does not converge
      in such a way that the Pad\'e-based approximation is unstable.
In these cases we use the $t \leftrightarrow u$ symmetry of the amplitude to provide an
alternative, stable, evaluation.
In practice, we also calculate all results with the variables
      $t$ and $u$ interchanged. For almost all generated data points,
      we observe perfect agreement with the results obtained without
      exploiting the $t$--$u$ symmetry. In the remaining cases, one of
      the results is clearly erroneous, manifesting as a distinct,
      spike-like deviation from the smooth curve
      which comes with a huge Pad\'e uncertainty.

%\clearpage

\begin{figure}[t]
%  \begin{tabular}{cc}
\centering
    \includegraphics[width=.45\textwidth]{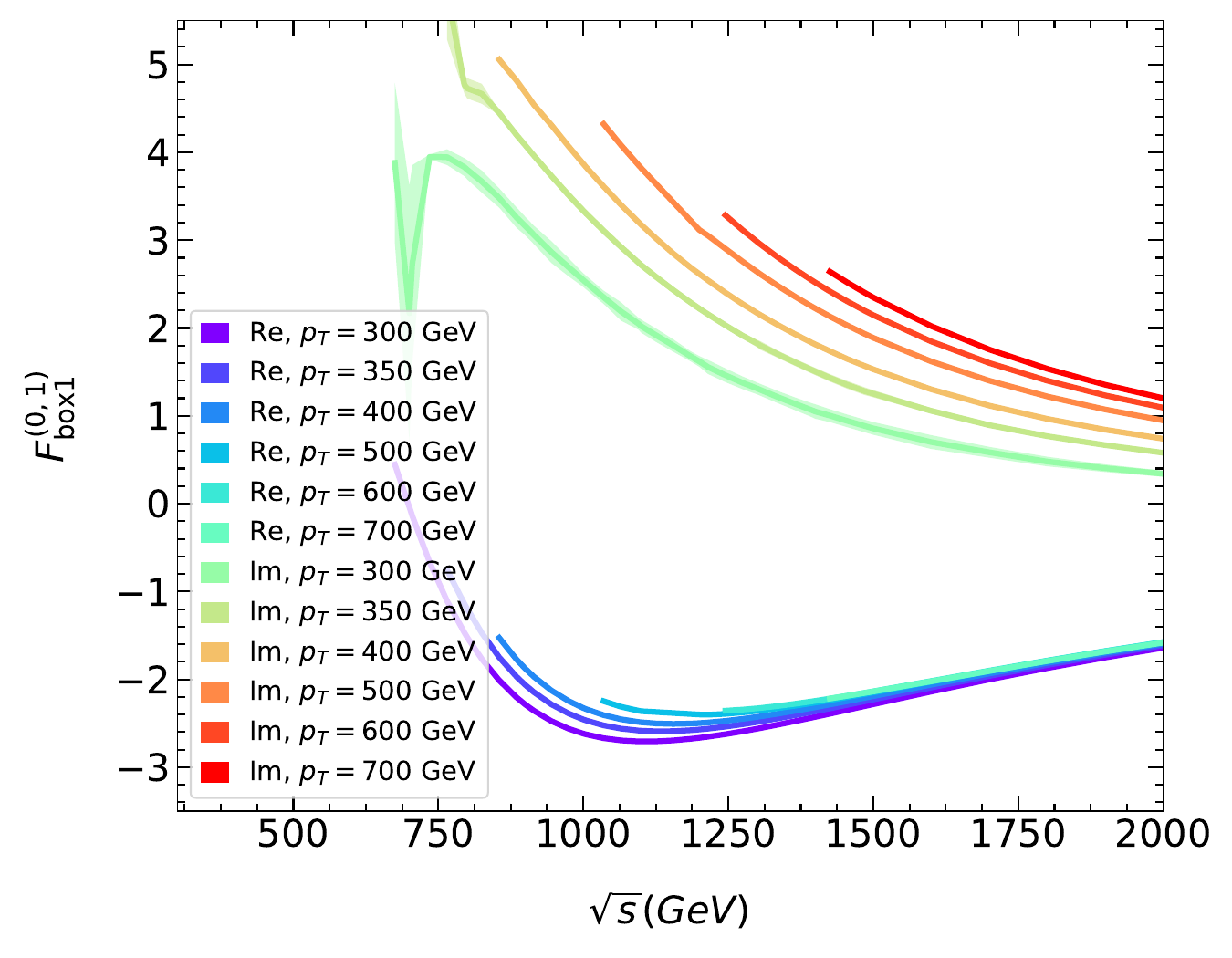}
    \includegraphics[width=.45\textwidth]{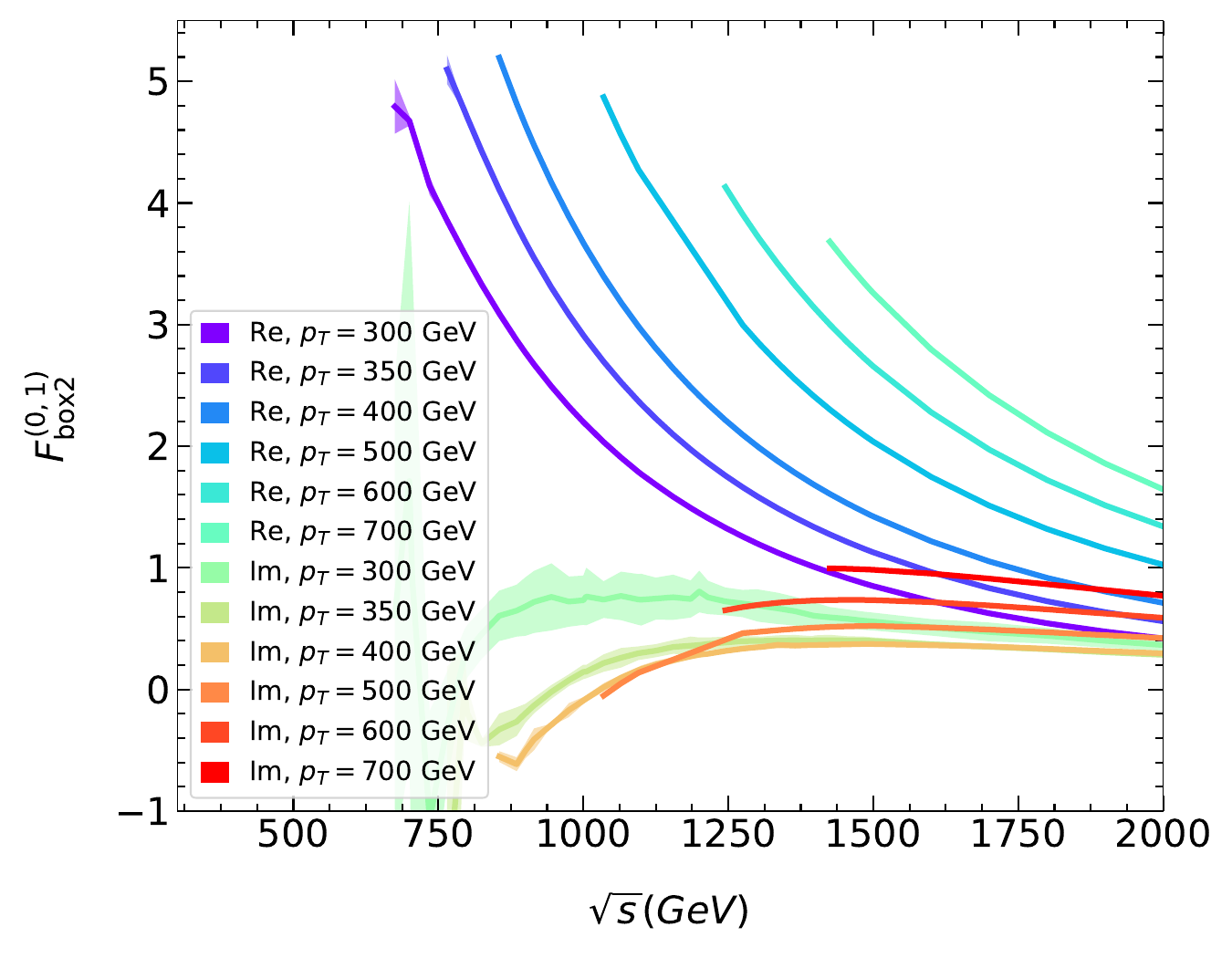}
    \\
    \includegraphics[width=.45\textwidth]{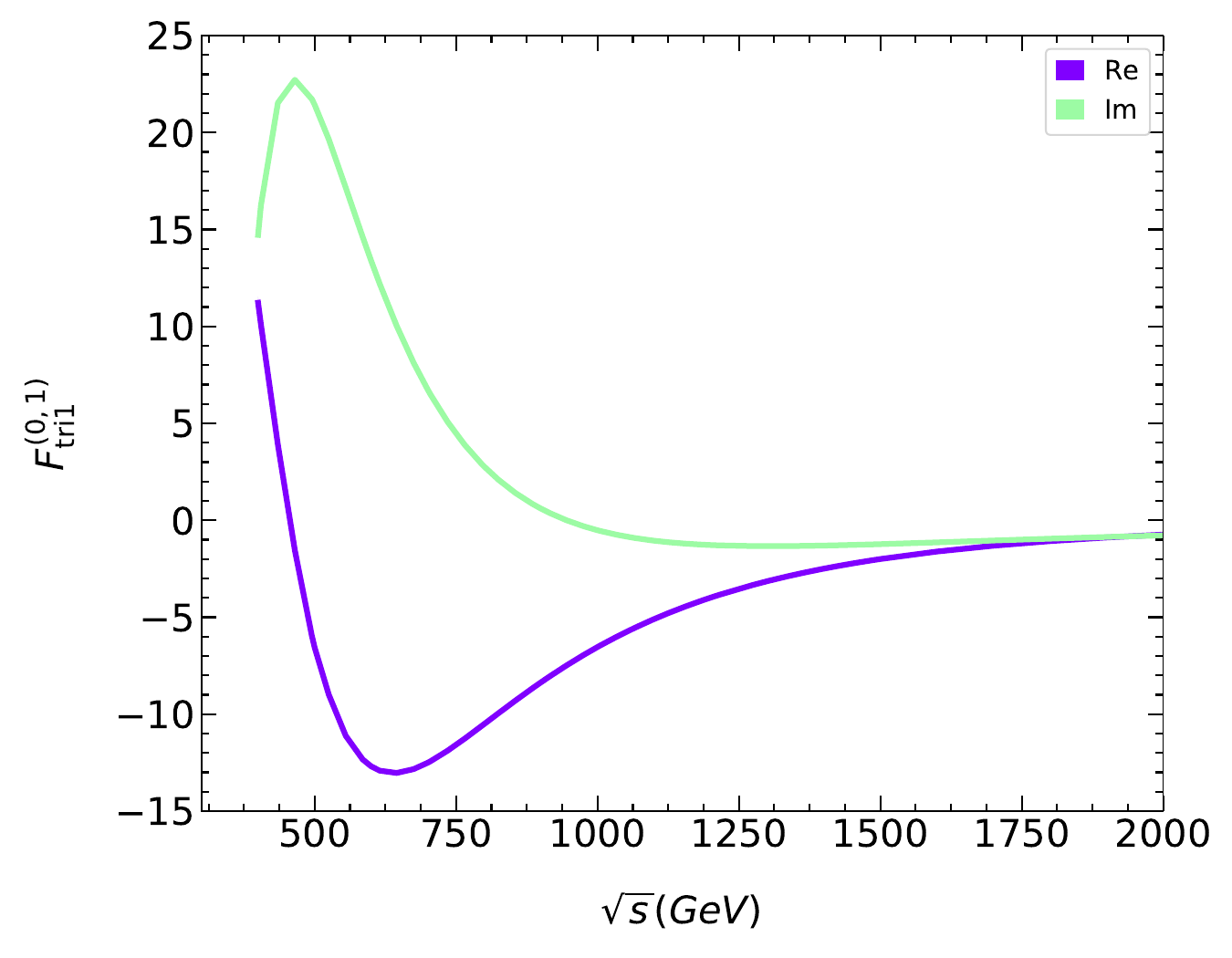}
%  \end{tabular}
  \caption{\label{fig::FF_he_pT}
  $F_{\rm box1}$, $F_{\rm box2}$ and $F_{\rm tri1}$ as a function of $\sqrt{s}$ for fixed $p_T$. Note that the triangle form factor
  is independent of $p_T$.
  }
\end{figure}

\begin{figure}[t]
  \begin{tabular}{cc}
    \includegraphics[width=.45\textwidth]{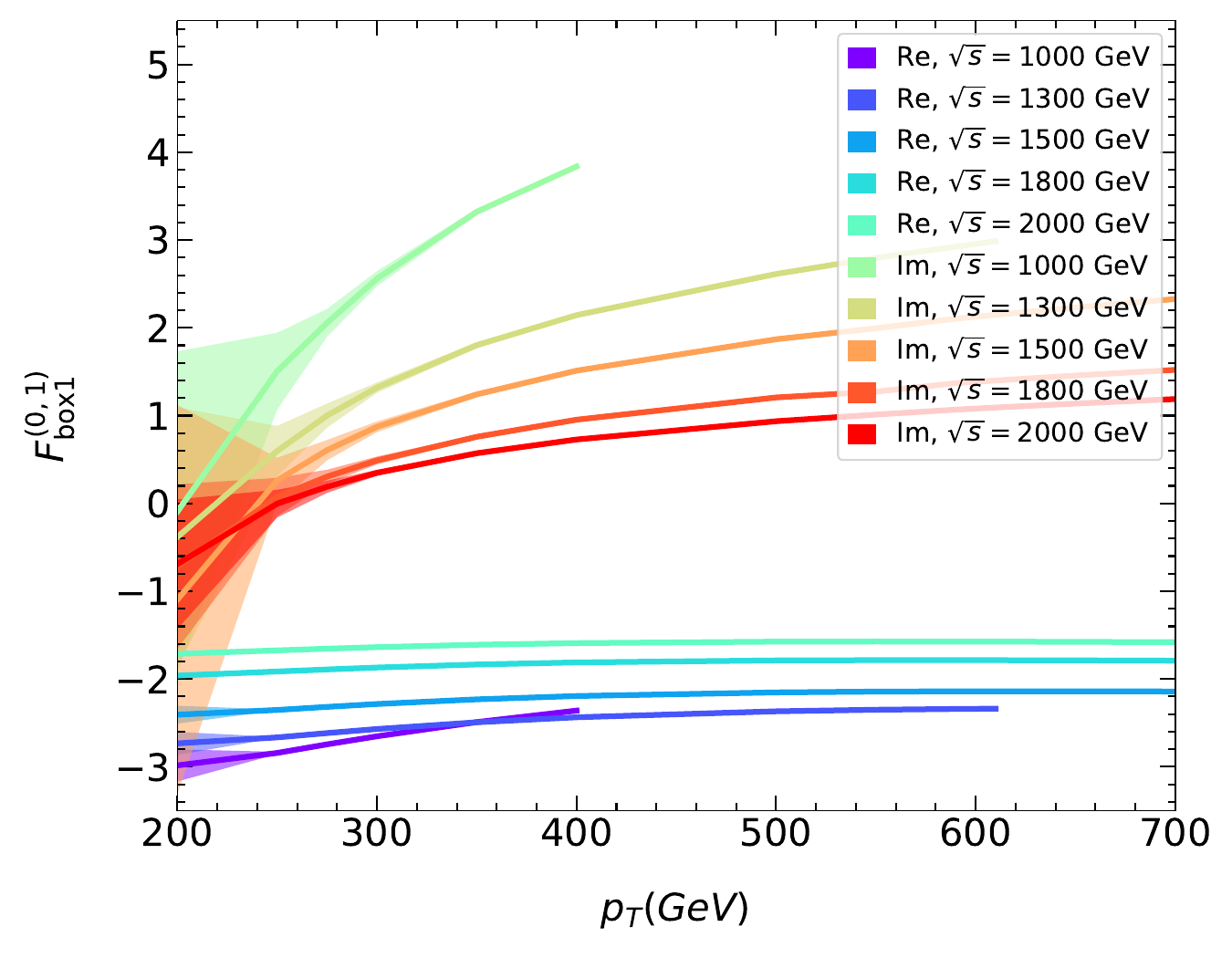}
    &
    \includegraphics[width=.45\textwidth]{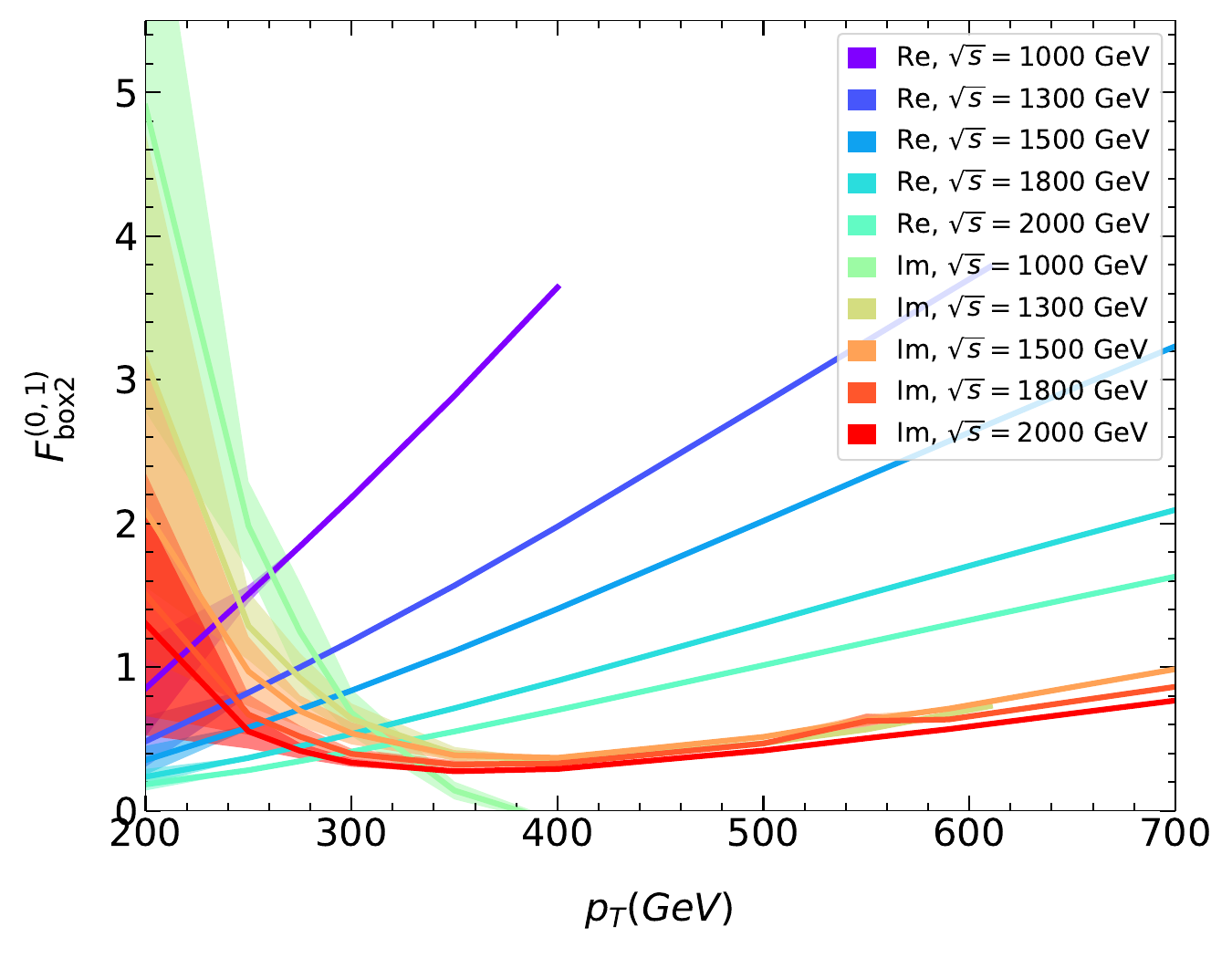}
  \end{tabular}
  \caption{\label{fig::FF_he_sqrts}
  $F_{\rm box1}$ and $F_{\rm box2}$ as a function of $p_T$ for fixed $\sqrt{s}$.
   }
\end{figure}

In Figs.~\ref{fig::FF_he_pT} and~\ref{fig::FF_he_sqrts} we show NLO form factors $F_{\rm tri1}^{(0,1)}$,
$F_{\rm box1}^{(0,1)}$ and $F_{\rm box2}^{(0,1)}$
for fixed $p_T$ or fixed $\sqrt{s}$, respectively. In all cases we show the highest available approximation 
which includes $\delta^4$ and $(m_H^{\rm ext})^4$
terms, together with
the corresponding uncertainty band
from the Pad\'e procedure.\footnote{The estimated $\pm2\%$ uncertainty from the truncation in $\delta$ and $m_H^{\rm ext}$ is not shown.} This band is
only visible for $p_T\lesssim 350$~GeV
and only for lower values of $\sqrt{s}$
as can be seen in Fig.~\ref{fig::FF_he_pT}.
It is, in general, more pronounced for the imaginary parts than for the real parts.
From Fig.~\ref{fig::FF_he_sqrts} we observe that
for $p_T\gtrsim350$~GeV the uncertainty induced
by the Pad\'e approximation is well 
below the uncertainty due to the truncated 
expansion in $\delta$ and $m_H$.

% \clearpage

\begin{figure}[t]
\centering
    \includegraphics[width=.75\textwidth]{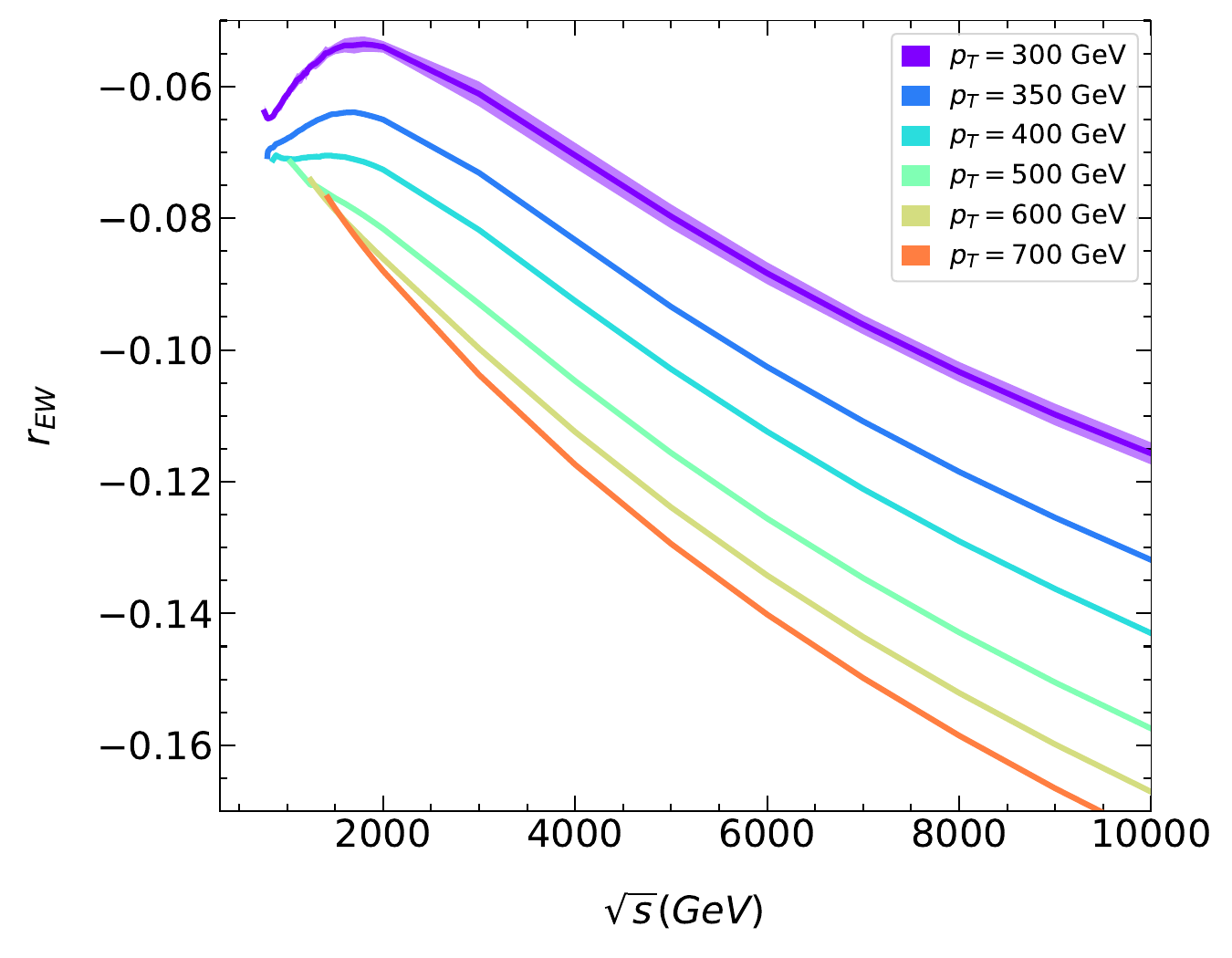}
  \caption{\label{fig::FF_he_ratio}
   $r_{\rm EW}$ for various values of $p_T$ as a function of $\sqrt{s}$.
   }
\end{figure}

In Fig.~\ref{fig::FF_he_ratio}, we show the ratio $r_{\rm EW}$
for fixed values of $p_T$ as a function of $\sqrt{s}$ which we extend up to 10~TeV.
For each curve we use the highest available approximation and 
show, besides the central value, also the uncertainty bands as
obtained from the Pad\'e approximation. These are only visible for
$p_T\lesssim 300$~GeV. 
For $p_T\ge 350$~GeV we obtain numerically stable results.
The result for $r_{\rm EW}$ shows a more stable behaviour than for the individual (box) form factors. This can be explained by
the dominance of the triangle form factors
for smaller value of $\sqrt{s}$.
For larger $\sqrt{s}$ the approximations for
the box form factors are stable for all
$p_T$ values shown in Fig.~\ref{fig::FF_he_ratio}.
The partonic quantity $r_{\rm EW}$
is negative and of the order of $-10\%$,
which is in 
agreement with the
observations from Ref.~\cite{Bi:2023bnq}
for the hadronic invariant mass or $p_T$ distribution.

%- }}}
%- {{{ Conclusions and outlook:
\FloatBarrier
%\newpage
\section{\label{sec::con}Conclusions and outlook}

In this work, we compute NLO electroweak corrections to Higgs boson pair
production in gluon fusion. We consider the top quark-induced contributions in
the high-energy limit. Since we include all sectors of the Standard Model,
several mass scales are involved in the computation.  We identify certain
hierarchies that allow us to perform nested expansions in several small
parameters.  In particular, we assume that the mass of the Higgs boson in the
final state is small as compared to the top quark mass. Furthermore, we expand
in mass differences in case several distinct masses appear inside the loop
diagrams. Most importantly, we perform a deep expansion with more than 100
expansion terms around the high-energy limit. 
The calculation of the electroweak corrections 
is much more challenging than in the QCD case; consequently, the intermediate and
final expressions are considerably larger.

From the convergence properties
of our results we estimate that the truncation uncertainty, dominated 
by the mass-difference expansion, is about $\pm2\%$
of the two-loop corrections,
which is well below the scale uncertainty from NLO QCD corrections.
A systematic improvement is possible by computing more terms
in the $\delta$ and $m_H^{\rm ext}$ expansion, which might be a
challenging task from the computational point of view.

In the limiting case of massless external Higgs bosons and small top-quark mass, the leading
high-energy expansion of the two-loop form factors 
starts at $(m_t^2/s)^0$,
while at one loop the leading term starts at $(m_t^2/s)^1$.
At NLO we observe a quadratic and cubic logarithm as the
leading contribution for $F_{\rm box1}$ and $F_{\rm box2}$,
respectively.

Our results are explicit in all input parameters, allowing for a
flexible use of the form factors and straightforward numerical evaluation. 
The results obtained in this paper cover the phase space for
$p_T\gtrsim350$~GeV
and lead to corrections of the order of $-10\%$ at the partonic level in the high-energy limit. 
The high-energy results serve as cross
check for other calculations, e.g., those based on numerical
approaches (see, e.g.,
Refs.~\cite{Bi:2023bnq,Heinrich:2024dnz}).  Furthermore,
it is expected that in combination with the forward limit expansion,
the entire phase space can be covered as in the case of QCD (see,
e.g., Ref.~\cite{Davies:2025qjr}).

Low-order expansions of our results are available from~\cite{progdata}, and
deep expansions are available upon request.  Our results will be made
public in the context of \texttt{ggxy}~\cite{Davies:2025qjr}, which provides a framework to compute
hadronic cross sections and distributions for gluon-induced processes.

%- }}}

%- {{{ Ackn.:

\section*{Acknowledgements}

This work is supported by the Deutsche Forschungsgemeinschaft (DFG, German
Research Foundation) under grant 396021762 --- TRR 257 ``Particle Physics
Phenomenology after the Higgs Discovery''.
The work of J.~D.~is supported by STFC Consolidated Grant ST/X000699/1. 
K.S. and H.Z.  are supported by the European Union under the Marie Sk{\l}odowska-Curie Actions (MSCA) Grants 101204018 and 101202083.
We thank Stefano Pozzorini for discussions and Marco Vitti for feedback on the manuscript.

%- }}}

\begin{appendix}

\section{Sample numerical results for the form factors}

For convenience we provide in this appendix
results for the form factors for a few representative phase-space points.
In Tab.~\ref{tab:benchmark}, we present numerical results for one-
and two-loop form factors, as defined in Eqs.~\eqref{eq::calM} and \eqref{eq::F}. We use the value for $\sqrt{s}$ and $p_T$
given in the table together with the
numerical values of input parameters are reported in Eq.~\eqref{eq::numvals}.
We  provide further detailed benchmark results by splitting the bare two-loop and counterterm contributions in the ancillary files to this paper~\cite{progdata}.

%%%%%%%%
\begin{table}[htb!]
    \centering \footnotesize
    \renewcommand{\arraystretch}{1.1}
    \begin{tabular}{c|c|c|c|c|c|c|c}
         $\sqrt{s}$ & $p_T$   & $F_{\rm tri1}^{(0)}$ & $F_{\rm box1}^{(0)}$ & $F_{\rm box2}^{(0)}$ & $F_{\rm tri1}^{(0,1)}$ & $F_{\rm box1}^{(0,1)}$ & $F_{\rm box2}^{(0,1)}$  \\ 
         $\rm [GeV]$ &  $\rm [GeV]$ & & & & & \\ \hline
         1000 & 350 & $0.0062605$ & $0.14270$  & $-0.025741$ & $-6.53488$ & $-2.45552(60)$ & $2.9101(27)$ \\
          & & $+ 0.054356 \, i$ & $-0.39457 \, i$ & $-0.17797\,i$ & $-0.51315 \, i$  & $+3.311(33) \, i$ & $+0.195(56)\,i$ \\ 
          \hline
          1000 & 400 & $0.0062605$ & $0.15179$ & $-0.028158$ & $-6.53488$ & $-2.32639(15)$ & $3.6713(34)$ \\ 
          && $+ 0.054356\, i$ & $-0.37594\,i$ & $-0.21204\,i$ & $-0.51315 \, i$ & $+3.8610(70)\,i$ & $-0.064(23)\,i$ \\ 
          \hline
          %%%%
          2000 & 350 & $-0.0016832$ & $0.056402$ & $-0.00085560$ & $ -0.73629$  & $-1.61342(72)$ & $0.5623(23)$ \\
          & & $+0.0052336\, i$ & $-0.074252\,i$ & $-0.040586\,i$ & $-0.77771 \, i $ & $+0.571(14)\,i$ & $+0.299(17)\,i$ \\ \hline
          2000 & 400   & $-0.0016832$ & $0.057281$ & $-0.00067097$ & $ -0.73629$ & $-1.5945(10)$ & $0.71184(40)$ \\ 
          & & $+0.0052336\, i$ & $-0.068858\,i$ & $-0.047603\,i$ & $-0.77771 \, i $ & $+0.7359(60)\,i$ & $+0.3011(74)\,i$\\ \hline
          2000 & 500  & $-0.0016832$ & $0.058756$ & $-0.000070987$ & $ -0.73629$ & $-1.57687(5)$ & $1.02555(3)$ \\
          & & $+0.0052336\, i$ & $-0.059640\,i$ & $-0.060279\,i$ & $-0.77771 \, i $ & $+0.9467(13)\,i$ & $+0.4226(11)\,i$\\ \hline
          2000 & 600  & $-0.0016832$ & $0.059907$ & $0.00071520$ & $ -0.73629$ & $-1.57578$ & $1.34029$ \\
          & & $+0.0052336\, i$ & $-0.052344\,i$ & $-0.071229\,i$ & $-0.77771 \, i $ & $+1.09237\,i$ & $+0.58935\,i$\\ \hline 
          2000 & 700  & $-0.0016832$ & $0.060806$ & $0.0015758$ & $ -0.73629$ & $-1.58221$ & $1.64503$ \\
          & & $+0.0052336\, i$ & $-0.046603\,i$ & $-0.080752\,i$ & $-0.77771 \, i $ & $+1.20077\,i$  & $+0.77052\,i$
    \end{tabular}
    \caption{Sample numerical results for one- and two-loop form factors. The results correspond to 
    on-shell renormalization is in the $G_\mu$ scheme, see also Eq.~\eqref{eq::alphaGmu}.
    For the Pad\'e approximation, we choose $\{N_{\rm low},N_{\rm high}\}=
\{45,54\}$. The uncertainties displayed in brackets, where visible, are from the Pad\'e procedure. We assign an additional $\pm 2\%$ uncertainty to the two-loop form factors to account for the truncation of 
the $\delta$ and $m_H^{\rm ext}$ expansions, see the discussion at the end of Section \ref{sec::numresults}.
 }
    \label{tab:benchmark}
\end{table}
%%%%%%%%%

\end{appendix}

%- {{{ Bibl.:

%\bibliographystyle{JHEP} 
\bibliographystyle{jhep}
%\footnotesize
\bibliography{inspire.bib,extra.bib}

%- }}}

\end{document}